\pdfoutput=1

\documentclass[twocolumn,superscriptaddress,aps,nofootinbib,groupedaddress]{revtex4-1}

\usepackage[utf8]{inputenc}
\usepackage{braket}
\usepackage{amsmath,amsthm,amssymb,amsfonts}
\usepackage[breaklinks=true,colorlinks=true,linkcolor=blue,urlcolor=blue,citecolor=blue,pagebackref=false]{hyperref}
\usepackage{graphicx}
\usepackage{color}
\usepackage{times}
\usepackage{tikz}
\usepackage{outlines}
\usepackage{siunitx}
\usepackage{upgreek}
\usepackage{soul}
\usepackage{color}
\usepackage[normalem]{ulem}

\newcommand\blfootnote[1]{%
  \begingroup
  \renewcommand\thefootnote{}\footnote{#1}%
  \addtocounter{footnote}{-1}%
  \endgroup
}

\date{\today}

\begin{document}
\title{
Long-range electron-electron interactions in quantum dot systems and applications in quantum chemistry}

\author{J. Kn\"orzer,$^{1,2,*,\dagger}$ C. J. van Diepen,$^{3,*}$ T.-K. Hsiao,$^{3}$ G. Giedke,$^{4,5}$ U. Mukhopadhyay,$^{3}$ C. Reichl,$^{6}$ W. Wegscheider,$^{6}$ J. I. Cirac,$^{1,2,\ddagger}$, and L. M. K. Vandersypen$^{3,\ddagger}$}
%% Affiliations
\affiliation{$^{1}$Max-Planck-Institut f\"ur Quantenoptik, Hans-Kopfermann-Str. 1, 85748 Garching, Germany}
\affiliation{$^{2}$Munich Center for Quantum Science and Technology, Schellingstr.~4, 80799 München, Germany}
\affiliation{$^{3}$QuTech and Kavli Institute of Nanoscience, Delft University of Technology, 2600 GA Delft, Netherlands}
\affiliation{$^{4}$Donostia International Physics Center, Paseo Manuel de Lardizabal 4, E-20018 San Sebasti\'{a}n, Spain}
\affiliation{$^{5}$Ikerbasque Foundation for Science, Maria Diaz de Haro 3, E-48013 Bilbao, Spain}
\affiliation{$^{6}$Solid State Physics Laboratory, ETH Z\"urich, 8093 Z\"urich, Switzerland}
\blfootnote{$^{*}$These authors contributed equally to this work.}
\blfootnote{$^{\dagger}$Present address: Institute for Theoretical Studies, ETH Zurich.}
\blfootnote{$^{\ddagger}$
ignacio.cirac@mpq.mpg.de;
l.m.k.vandersypen@tudelft.nl}

\begin{abstract}
Long-range interactions play a key role in several phenomena of quantum physics and chemistry.
To study these phenomena, analog quantum simulators provide an appealing alternative to classical numerical methods.
Gate-defined quantum dots have been established as a platform for quantum simulation, but for those experiments the effect of long-range interactions between the electrons did not play a crucial role.
Here we present the first detailed experimental characterization of long-range electron-electron interactions in an array of gate-defined semiconductor quantum dots.
We demonstrate significant interaction strength among electrons that are separated by up to four sites, and show that our theoretical prediction of the screening effects matches well the experimental results.
Based on these findings, we investigate how long-range interactions in quantum-dot arrays may be utilized for analog simulations of artificial quantum matter.
We numerically show that about ten quantum dots are sufficient to observe binding for a one-dimensional H${}_2$-like molecule.
These combined experimental and theoretical results pave the way for future quantum simulations with quantum dot arrays and benchmarks of numerical methods in quantum chemistry.
\end{abstract}
\maketitle

\section{Introduction}
Electromagnetic forces between electrons play a crucial role in quantum physics and chemistry.
They are a key ingredient for many phenomena, ranging from molecular binding~\cite{French2010}, Wigner crystallization~\cite{Wigner1934}, and exciton formation~\cite{Frenkel1931} to high-temperature superconductivity~\cite{Anderson2013}.
While their exact treatment in quantum many-body systems remains numerically challenging, analog quantum simulators \cite{Cirac2012, Georgescu2014} offer an alternative setting for the study of complex quantum systems with long-range interactions.
Yet in most physical systems, particles interact locally or at short distances only.
It is therefore of great interest to investigate experimental platforms in which long-range interactions between charged particles occur naturally.

Electrons confined to semiconductor quantum dots (QDs) provide a versatile test bed for analog quantum simulation of Fermi-Hubbard physics~\cite{Manousakis2002, Byrnes2008, Georgescu2014}.
Previous experimental studies have addressed the transition from Coulomb blockade to collective Coulomb blockade~\cite{Hensgens2017}, itinerant ferromagnetism when doping with a single hole~\cite{Dehollain2020}, and Heisenberg magnetism arising in the Mott-insulator regime~\cite{VanDiepen2021b}.
However long-range electron-electron interactions have not yet been capitalized on in gate-defined QD systems, and their effect was previously either tuned away or remained as an unwanted disturbance. 
The character of the electron-electron interaction is also relevant for the operation of spin qubits in QD arrays with shared control lines~\cite{Li2018}.

Transport measurements on double quantum dots (DQDs) more than two decades ago~\cite{Wiel2003} already revealed interactions between electrons on neighbouring sites. 
In more recent work, these interactions have been used to induce entanglement between spin qubits in separate DQDs~\cite{Shulman2012}.
Interactions between electrons separated by multiple sites have been studied to assess the readout performance~\cite{Zajac2016}, the tunability of the interaction between DQD qubits~\cite{Neyens2019} and
to induce a cascade of electrons enabling distant spin readout~\cite{VanDiepen2021}.
However, a detailed study of electron-electron interactions as a function of distance has not yet been performed. 
While such a characterization is technically challenging as it requires a high degree of control over the potential landscape of a sufficiently large system, recent achievements in tuning and controlling multi-dot arrays \cite{Volk2019,Mills2019,Hsiao2020,Qiao2020} facilitate the formation of increasingly large and homogeneously coupled QD arrays. 

In this work, we present the first detailed examination of long-range electrostatic interactions between electrons confined to a semiconductor QD system. In our study we operate a six-site QD array with homogeneous tunnel coupling and two charge sensors.
We record a charge-stability diagram for each dot-dot pair in the array, thus explicitly accounting for non-nearest neighbour interactions.
From the analysis of these diagrams we extract the electron-electron interaction potential as a function of distance, which is shown and discussed in Sec.~\ref{ssec:characterization-interaction}.
We detect interactions between electrons that are up to four sites away from each other.
Furthermore, we model the interaction numerically, taking screening effects due to metallic gates into account, and find good agreement between experiment and theory.
As a promising application of our findings, we discuss prospects for analog simulations of low-dimensional, artificial atoms and molecules in Sec.~\ref{sec:qchem}, that may help to benchmark and improve existing numerical methods in quantum chemistry (QC).
Inspired by the ideas outlined in a recent proposal for neutral atoms in optical lattices~\cite{Arguello2019}, we start from the tight-binding description of our QD system and regard it as a linear discretization of an artificial atom, or molecule.
We calculate the low-lying eigenstates of the tight-binding model and discuss their relation to the simulated chemical systems.
Based on our numerical results, we project that QD arrays with $\approx 10$ sites are sufficient for proof-of-principle simulations of molecular dissociation, which is within reach of state-of-the-art experiments.
We also discuss relevant experimental techniques for the implementation of QC with QD arrays. 
Finally, in Sec.~\ref{sec:outlook}, we summarize our findings and give perspectives for future work.

\section{Characterization of interaction potential \label{ssec:characterization-interaction}}
In this section, we introduce the experimental system and its theoretical model, which is employed in the later analysis.
Subsequently we present the experimental characterization of on-site and inter-site interactions in the QD array, and show that these agree well with results from numerical calculations.

\begin{figure}[b]
   \centering
    \includegraphics[width=1.0\columnwidth]{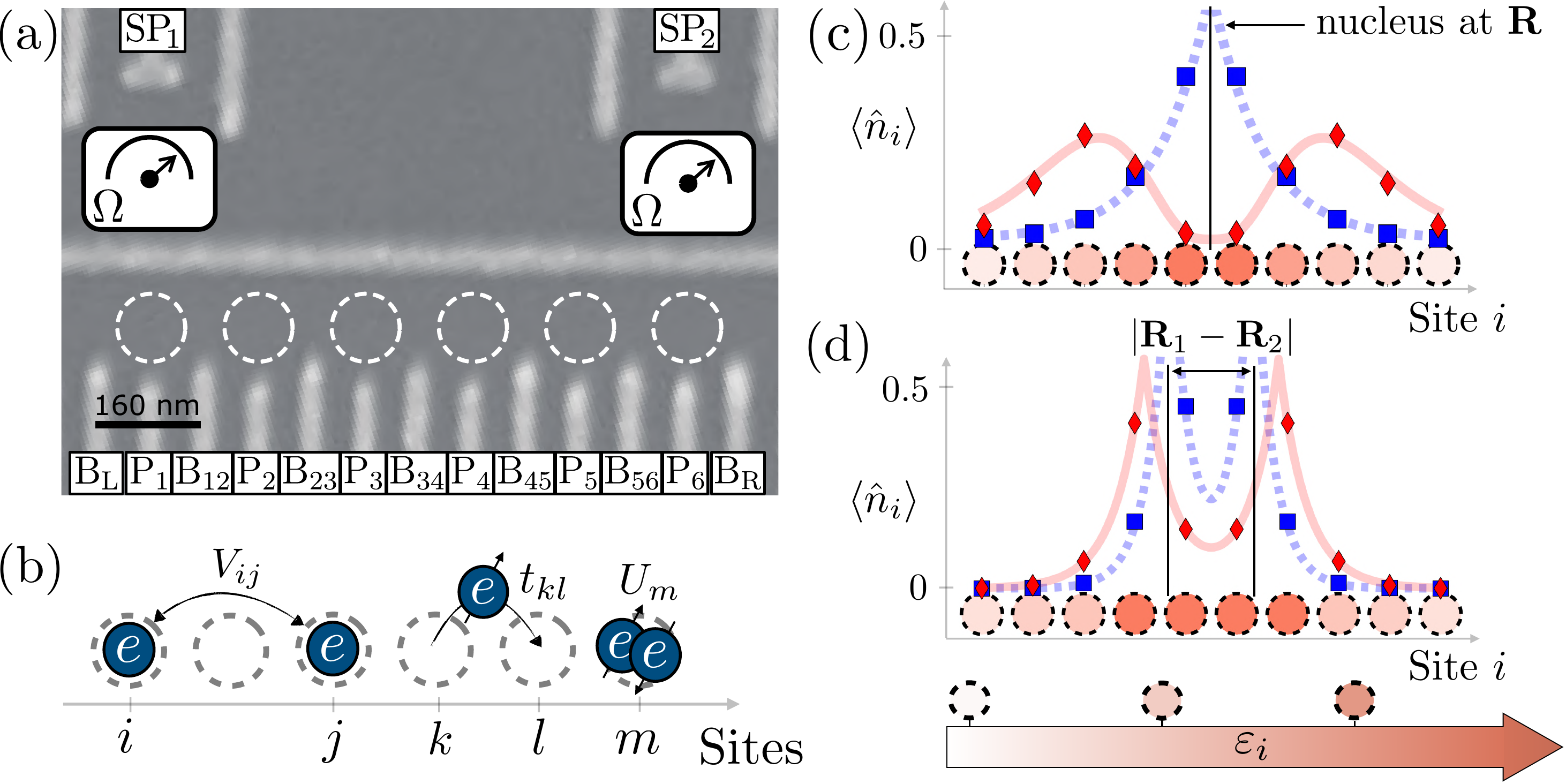}
   \caption{
   Quantum dot system with schematics of chemistry simulations.
   (a) Scanning electron micrograph of the device.
   Dot locations are indicated with dashed circles, and charge sensors by resistance meters labeled with $\Omega$.
   Barrier gates between sites $i$ and $j$ are labeled by $\mathrm{B_{ij}}$, and plunger gates at site $k$ are labeled by $\mathrm{P_k}$.
   Charge sensors are controlled by gates $\mathrm{SP_1}$ and $\mathrm{SP_2}$, respectively.
   (b) Sketch of competing terms in Eq.~\eqref{eq:fermi-hubbard} describing the quantum-dot array:
   inter-site interaction $V_{ij}$, tunnel coupling $t_{kl}$, and on-site interaction $U_m$.
   (c) Schematic of QD array and its relation to atomic QC simulation, exemplarily shown for $10$ QDs.
   The electron-nucleus interaction potential of an artificial hydrogen atom is encoded in the local energy offsets $\varepsilon_i$.
   (d) Similar setup in case of the molecular ion $H_2^+$ with two nuclei at $\mathbf{R}_1$ and $\mathbf{R}_2$, respectively.
   The system can be studied for different internuclear distances $|\mathbf{R}_1-\mathbf{R}_2|$ separately.
    Blue squares (red diamonds) indicate the local expectation values $\langle \hat n_i \rangle$ in the ground state (first excited state), and blue-dashed (red-solid) line shows fit to ground-state (excited-state) wavefunction \cite{Loudon2016,Loos2015}. The orange colorscale encodes the local potential offset.
   }
   \label{fig:schematic}
\end{figure}

\textit{Device}.\textemdash
The QD system consists of a linear array of six QDs and two charge sensors, and is formed in a GaAs/AlGaAs heterostructure. Figure~\ref{fig:schematic}(a) shows a scanning electron micrograph image of the active region of a device similar to the one used in this experiment, which is designed for up to eight QDs with two charge sensors and has previously been operated as a Heisenberg spin chain~\cite{VanDiepen2021b}. At the GaAs/AlGaAs interface, \SI{90}{\nano\meter} below the surface and \SI{40}{\nano\meter} below a silicon doping layer, a two-dimensional electron gas forms.
The potential landscape at the interface is shaped by applying voltages on the gates, which are patterned at the surface with a dot spacing $a_{\mathrm{QD}}= 160~\mathrm{nm}$.
The device is cooled in a dilution refrigerator, which results in an electron reservoir temperature of about \SI{100}{\milli\kelvin} (roughly \SI{10}{\micro\electronvolt}).

\textit{Tight-binding model.}\textemdash
To characterize the interaction potential between electrons, we start from a tight-binding description and consider the single-band extended Fermi-Hubbard model~\cite{Imada1998, Yang2011, Hensgens2017}
\begin{equation}\label{eq:fermi-hubbard}
\begin{aligned}
H =& \underset{=: H_\mathrm{ee}}{\underbrace{\sum_i U_i n_{i\uparrow} n_{i\downarrow} + \sum_{i\neq j} V_{ij} n_i n_j}} \underset{=: H_\mathrm{ne}}{\underbrace{- \sum_i \varepsilon_i n_i}} \\
&\underset{=: H_\mathrm{kin}}{\underbrace{- \sum_{\langle i, j \rangle,\sigma} t_{ij}  c^{\dagger}_{i\sigma} c_{j\sigma}}},
\end{aligned}
\end{equation}
where the dot occupation at site $i$ is denoted by $n_i = n_{i\uparrow} + n_{i\downarrow}$ and $n_{i\sigma}= c_{i\sigma}^{\dagger} c_{i\sigma}$ with the annihilation (creation) operator $c^{(\dagger)}_{i\sigma}$ for an electron with spin $\sigma$.
The last sum is restricted to nearest-neighbour hopping only, as indicated by $\langle \cdot, \cdot\rangle$.
$t_{ij}$ denotes the tunnel coupling between sites $i$ and $j$, $\varepsilon_i$ the local energy offset, $U_i$ the on-site interaction and $V_{ij}$ is the inter-site interaction strength, see also Fig.~\ref{fig:schematic}(b).
The local energy offsets, $\varepsilon_i$, and the tunnel couplings, $t_{ij}$, can be tuned with the gate voltages, thus can be adjusted to the problem under study.
The Hamiltonian consists of a contribution from electronic interactions ($H_\mathrm{ee}$), potential ($H_\mathrm{ne}$), and electron hopping ($H_\mathrm{kin}$). This notation will prove useful for the discussion in Sec.~\ref{sec:qchem}.

\begin{figure*}
   \centering
    \includegraphics[width=0.97\textwidth]{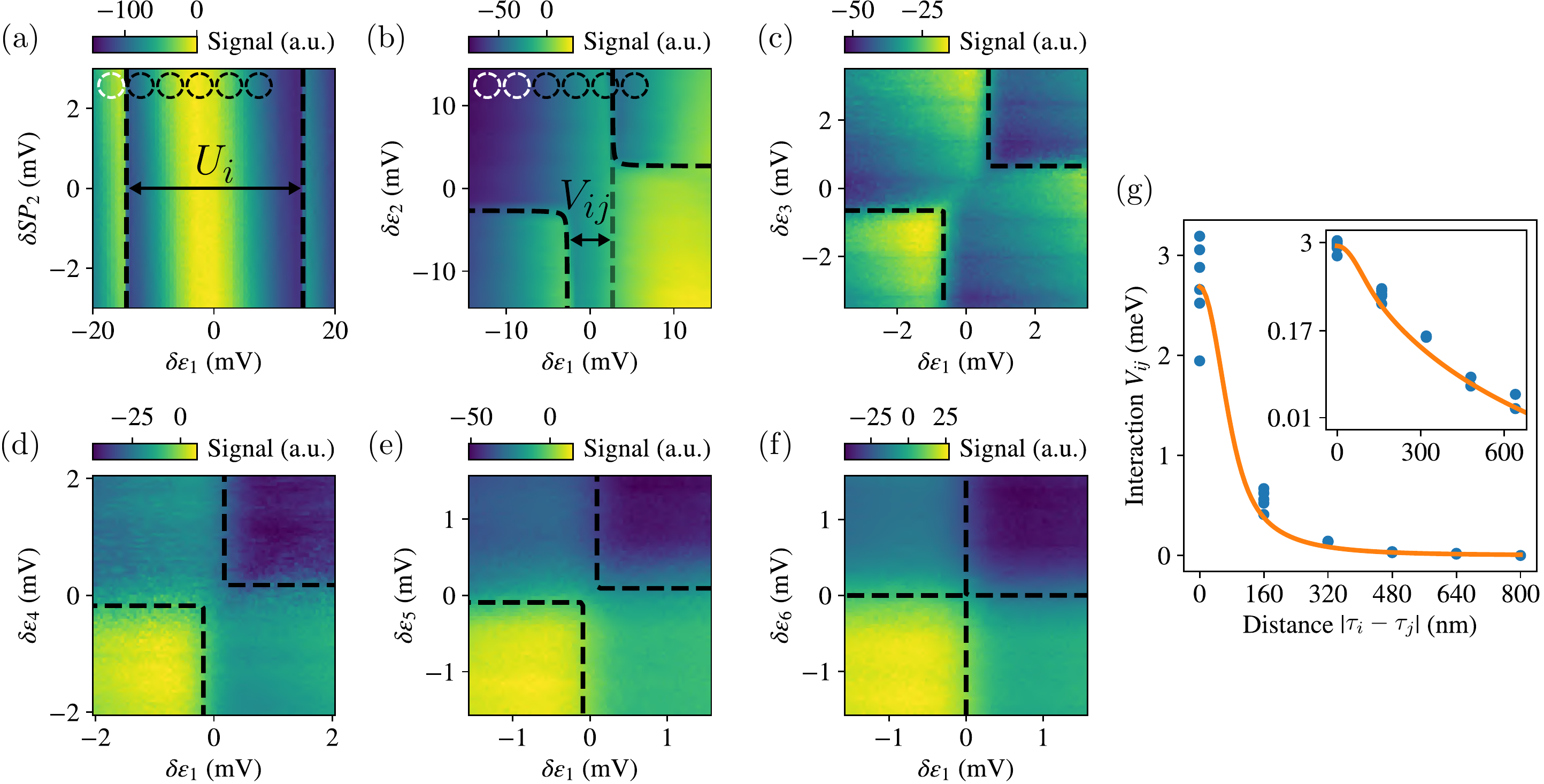}
   \caption{(a)-(f) Charge-stability diagrams showing the sum of signals from both charge sensors as function of the local energy offsets for the dots highlighted by the white dashed circles. (a) For the leftmost dot only the left sensor signal is shown, and is measured relative to the gate for the rightmost sensor. This gate here acts as a dummy, because it has negligible effect on both the leftmost dot and the signal from the left sensor. The broad vertical band is a Coulomb peak for the left sensor, which appears due to crosstalk from $\varepsilon_1$. For the leftmost dot in combination with a dot (b) one, (c) two, (d) three, (e) four, and (f) five sites away. Black dashed lines are fits to the anti-crossings, and are used to extract the interaction elements; the lighter dashed line in (b) is a guide to the eye to indicate the voltage of the shifted vertical addition line. The voltages are converted to energies with the lever arms $\{105, 94, 104, 86, 104, 95\} \hspace{2pt} \si{\micro\electronvolt\per\milli\volt}$, which were obtained with photon-assisted tunnelling experiments~\cite{VanDiepen2018}. (g) Interaction matrix elements $V_{ij}$ versus distance between QDs $i$ and $j$, which are centered around $\mathbf{\tau}_i$ and $\mathbf{\tau}_j$, respectively.
   Blue dots indicate the experimentally obtained interaction elements and the solid line shows the numerical result based on screening due to the gate metal as described in detail in App.~\ref{app:screening}. The uncertainties in the interaction elements are dominated by uncertainties in the lever arms, estimated to be below 10\%, and the electron temperature ($\approx \SI{10}{\micro\electronvolt}$).
   The inset shows the interaction and fits on a logarithmic scale for better comparison. 
   For simplicity $U_i$ is denoted as $V_{ii}$.}
   \label{fig:int_exp}
\end{figure*}

\textit{Experimental results}.\textemdash
The interaction potential is characterized based on a set of charge-stability diagrams.
This set contains a diagram for each dot and a diagram for each pair of dots. 
Before the diagrams are measured, the voltages are tuned to the center of the charge region with one electron per dot and homogeneous nearest-neighbour tunnel couplings, $t\approx\SI{20}{\micro\electronvolt}$. 
From that configuration, for each diagram only the chemical potentials for the respective dots are temporarily changed.
For the pairwise diagrams the respective chemical potentials are, in addition, offset to center at an anti-crossing where a charge is added on one dot in the pair and removed from the other.
The diagrams involving the leftmost dot are shown in Fig.~\ref{fig:int_exp}(a)-(f).
The full set of pairwise diagrams, which for six QDs consists of fifteen diagrams, is shown in Fig.~\ref{fig:csd_full} in the appendix.
This set of diagrams can explicitly reveal crosstalk for the control of chemical potentials, thus also for non-nearest neighbour sites. Here this crosstalk has already been compensated for with virtual gates~\cite{Nowack2011, Hensgens2017, Volk2019, Mills2019}.

The on-site interaction elements, $U_i$, are obtained from the separation between the addition lines for the first and second electron on the respective dot, see Fig.~\ref{fig:int_exp}(a). 
The inter-site interaction elements, $V_{ij}$, are extracted by modeling the anti-crossings in the charge-stability diagrams with~\cite{Hensgens2017}
\begin{equation}\label{eq:vij-extraction}
\delta \varepsilon_i + \delta \varepsilon_j = \pm \left( V_{ij} + \sqrt{(\delta \varepsilon_i - \delta\varepsilon_j)^2 + 4 t_{ij}^2} \right),
\end{equation}
with $\delta \varepsilon_i = \varepsilon_i - \varepsilon_{i,0}$ where $\varepsilon_{i,0}$ is the local energy offset at the center of the respective anti-crossing.
The anti-crossing model is converted into a two-dimensional patch, which is fitted onto the charge-stability diagrams, see Fig.~\ref{fig:int_exp}(b)-(f), using an edge detection algorithm~\cite{2015opencv}. 
Figure \ref{fig:int_exp}(g) shows all values for $U_i$ and $V_{ij}$ as extracted from the full set of diagrams.
The interaction strength shows a clear decay with distance, and is significant up to a distance of four sites.
The spread in interaction values for a fixed distance is explained by residual disorder in the potential landscape, which most noticeably affects the on-site interaction as it strongly depends on size and shape of the QD confinement.
A comparison of the inter-site interaction values between the left and right sides of QDs two to five did not reveal asymmetries, which indicates that the QDs are centered around the intended locations.

\textit{Numerical results.}\textemdash
The dominant source for screening of the interaction is the metal of the gates above the QDs, see Fig.~\ref{fig:schematic}(a). 
Other sources could contribute to screening, such as the surrounding two-dimensional electron gas, dopants and impurities, but are expected to be less important due to the device geometry or a relatively low density of mobile charge carriers.
The electrons on the QDs themselves are not expected to contribute to screening much because they are rather strongly confined to their respective QDs and were kept deep in the Coulomb blockade regime for the characterization of the interaction potential.
Based on this assumption, the depth of the interface and the gate pattern, we numerically calculate the screened interaction between two electrons as a function of distance.
It can be evaluated from the charge distribution induced in the metallic surface layer \cite{Segal2004}.
The underlying numerical approach is summarized in App.~\ref{app:screening}.
We compare the numerical results with the experimental results and observe good agreement.
In our calculations, we account for the finite dot size using a Gaussian basis set to describe the electronic wavefunctions, which yields a decay of $V_{ij}$ as shown in Fig.~\ref{fig:int_exp}(g).
Most notably we find that the electron-electron interaction potential $V_\mathrm{ee}$ decays polynomially with distance.
At large distance, our prediction is consistent with a subexponential (power-law) decay $V_\mathrm{ee}(|\mathbf{r}-\mathbf{r}^\prime|) \sim C/|\mathbf{r}-\mathbf{r}^\prime|^\alpha$ where $\alpha \lesssim 3$.
This finding is in agreement with an analytical estimate, based on the image-charge method, for a device whose surface is completely covered by metal and the fact that the present device is only partially covered by metallic gates, cf.~App.~\ref{app:screening}.
That estimate yields a cubic decay of interactions with distance, i.e., $\alpha = 3$, and more closely describes accumulation-mode devices with a multi-layer gate stack, such as those based on silicon and germanium~\cite{Zajac2016, Lawrie2020}.

\section{Applications in quantum chemistry \label{sec:qchem}}
In Sec.~\ref{ssec:characterization-interaction}, we have shown that electrons in the QD array interact at long distances via a screened Coulomb repulsion.
In the following, we study how this insight may be utilized in future experiments for the analog simulation of artificial atoms and molecules.

\subsection{Theoretical Framework \label{sec:theoretical-framework}}
A major task in QC is to study the low-energy physics of systems with $N_\mathrm{e}$ electrons and $N_\mathrm{c}$ nuclei.
Within the usually employed Born-Oppenheimer approximation, the positions of the nuclei are fixed at $\mathbf{R}_1, …, \mathbf{R}_{N_c}$, and the total Hamiltonian can be decomposed as $H_\mathrm{qc} = H_\mathrm{kin} + H_{ee} + H_\mathrm{ne}$ with a kinetic part, electron-electron and nucleus-electron interactions, respectively.
Here we consider a discretized instance of this problem, that can be simulated using a QD array with $N$ sites, cf.~Fig.~\ref{fig:schematic}.
We investigate a one-dimensional array as in Sec.~\ref{ssec:characterization-interaction},
but we note that the upcoming theoretical analysis can be extended to two-dimensional lattices.

We start from the model in Eq.~\eqref{eq:fermi-hubbard} with a homogeneous tunnel coupling $t$.
The kinetic term $H_\mathrm{kin}$ describes electrons hopping at a rate $t$, that are confined to an electrostatic potential landscape.
With the choice for the energy offsets
\begin{equation}\label{eq:epsilon_i}
\varepsilon_i = \sum_{k=1}^{N_c} V_\mathrm{ne}(|\mathbf{\tau}_i - \mathbf{R}_k|),
\end{equation}
we interpret $H_\mathrm{ne}$ from Eq.~\eqref{eq:fermi-hubbard} as an analog of the attractive nucleus-electron interaction in QC, where $\mathbf{\tau}_i$ denotes the location of the $i$th dot.
In the QD array, the form of the interaction potential $V_\mathrm{ne}$ may be adjusted by an adequate choice of chemical potentials.
For example, $\varepsilon_i = V_0/|\mathbf{\tau}_i-\mathbf{R}|$ may be chosen to mirror the Coulomb law in the presence of a single nucleus at position $\mathbf{R}$, with an interaction strength $V_0$.
Finally, the electron-electron interaction is captured by $H_\mathrm{ee}$ and has been characterized in Sec.~\ref{ssec:characterization-interaction}.
It is dictated by the actual interaction potential between charge carriers in the semiconductor.
While the exact form of $V_\mathrm{ee}$ [cf.~Eq.~\eqref{eq:matrix-elements}] differs from the behaviour of electrons in natural systems, it still permits long-distance interactions and allows for the simulation of artificial multi-electron atoms and molecules.
It is partly the presence of such long-range interactions that poses serious computational challenges in the numerical treatment of quantum matter, e.g., in molecular simulations.

Upon this reinterpretation of terms in the Fermi-Hubbard model \eqref{eq:fermi-hubbard}, the physical parameters of the QD array can be associated with the characteristic length and energy scales of the QC Hamiltonian.
In atomic physics, the effective Bohr radius $a_0$ relates the kinetic and potential energy scales.
In our system, these are characterized by nearest-neighbour hopping, $t$, and nearest-neighbour Coulomb interaction, $V_0/a_\mathrm{QD}$, respectively.
This enables us to identify $a_0 = t / (V_0/a_\mathrm{QD})$ and thus to introduce \cite{Arguello2020}
\begin{equation}\label{eq:qchem-parameters}
\eta \equiv \frac{a_0}{a_\mathrm{QD}} = \frac{t}{V_0}, \ \mathrm{Ry} = \frac{V_0^2}{t} = \eta^{-2} t, \\
\end{equation}
with the Rydberg energy $\mathrm{Ry}$.
The ratio $\eta$ in Eq.~\eqref{eq:qchem-parameters} determines the discretization error introduced by the finite array and relates to how well the continuum limit can be recovered.
In particular, $(\textit{i})$ at too small $\eta$ the atomic orbitals cannot be well-resolved, as for this the effective Bohr radius should span several QD sites, and $(\textit{ii})$ too large $\eta$ implies that the simulated atom or molecule does not fit into the array.

\subsection{Numerical results \label{sec:numerical-results}}
In the following we demonstrate that relatively small QD arrays can be used to study basic properties of simple artificial atoms and molecules.
For this aim, we calculate the low-energy eigenstates of the tight-binding Hamiltonian, see Eq.~\eqref{eq:fermi-hubbard}, for realistic system parameters and discuss finite-size effects. 

\begin{figure}[b]
   \centering
    \includegraphics[width=\columnwidth]{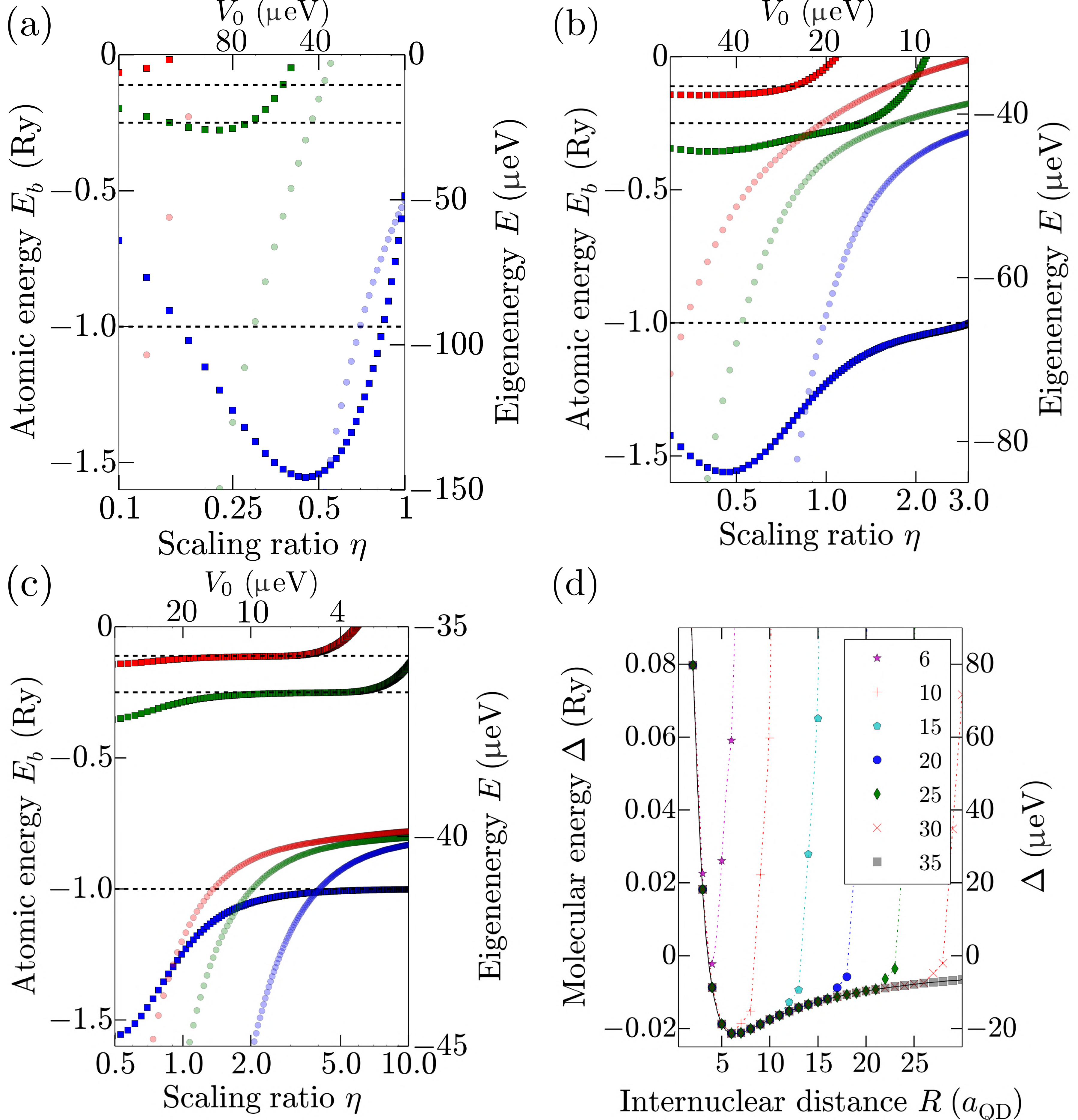}
   \caption{
   Numerical results for atom and two-electron molecule.
   Lowest part of artificial hydrogen-like atom spectrum as function of $V_0$ and $\eta = t/V_0$ for system sizes (a) $N=6$, (b) $N=25$ and (c) $N=300$.
   The left axis (\textit{dark squares}) shows the atomic binding energy $E_b$ in $\mathrm{Ry}$, while the right axis (\textit{light circles}) shows the corresponding eigenenergies of Eq.~\eqref{eq:fermi-hubbard} in $\upmu$eV.
   Shown are the ground (\textit{blue}), first-excited (\textit{green}) and second-excited (\textit{red}) state energies. 
   The dashed lines indicate the three lowest Balmer series values.
   In all cases, $t=20~\upmu$eV.
   (d) Molecular binding energy $\Delta$ (see text) for artificial $H_2$-like molecule with two electrons interacting via effective potential $V_\mathrm{ee}$, and two nuclei separated by distance $R$.
   Shown is the molecular binding energy for different system sizes $N = 6, 10, 15, ..., 35$.
   Curve fits to guide the eye (solid line for $N=35$ and dash-dotted lines for all other $N$).
   Other parameters: $t = 40~\upmu$eV, $V_0 = 200~\upmu$eV and $\mathrm{Ry} = 1~$meV.
   }
   \label{fig:chemistry}
\end{figure}

\textit{Atoms}.\textemdash
First we study an artificial hydrogen atom by setting $\varepsilon_i = V_0/|\mathbf{\tau}_i-\mathbf{R}|$ to define a single nucleus located at $\mathbf{R}$ in the center of the array, cf.~Fig.~\ref{fig:schematic}(c).
We calculate the low-energy spectrum of the system described by $H$ in Eq.~\eqref{eq:fermi-hubbard}, and relate the eigenenergy $E$ to an atomic binding energy $E_b = E + 2t$.
At a fixed tunnel coupling $t$, $V_0 = t/\eta$ needs to be optimized in order to approximate the continuum limit well.
With system size, $N$, sufficiently large, the Balmer-like series $E_n = -\mathrm{Ry}/n^2$ of a one-dimensional hydrogen atom \cite{Loudon2016} is reproduced if $\eta$ is neither too small nor too large.
Importantly, it can be seen that even relatively small arrays resolve quantized energy levels at intermediate $\eta$ and approximately yield the analytical result of the continuum case, see Figs.~\ref{fig:chemistry}(a)-(b).
As comparison see Fig.~\ref{fig:chemistry}(c) for a very large $N = 300$, where the energy plateaus coincide with the analytical result over a relatively wide range of $\eta$.
Signatures of this behaviour can already be observed at much smaller $N$, which shows that basic atomic properties can be studied using currently available setups with $\approx 10$ dots.
The probability density of the one-dimensional hydrogen atom is directly related to the ground and excited-state occupation numbers $\langle \hat n_i \rangle$, as depicted in Fig.~\ref{fig:schematic}(c).

\textit{Molecules}.\textemdash
Next we demonstrate the simulator's ability to uncover essential molecular properties.
For this aim we consider a system composed of two electrons and two nuclei.
In order to obtain the dissociation curve of this artificial $H_2$-like molecule, we calculate the low-energy spectrum of $H + V_\mathrm{nn}$, where $V_\mathrm{nn}$ denotes the nucleus-nucleus interaction potential.
The electron-electron interaction is governed by the screened interaction in the semiconductor, while for the electron-nucleus and nucleus-nucleus interactions, we consider a Coulomb potential, i.e., $V_\mathrm{nn} = V_0/R$.
The internuclear distance $R$ can be varied by adjusting the local offsets $\varepsilon_i$. 
Within the Born-Oppenheimer approximation, the positions of the nuclei are considered to be fixed, and thus the term $V_\mathrm{nn}$ is simply added to the measurement result $\braket{H}$ in the end.

We note that the interaction potential $V_\mathrm{ee}$ is device-specific and cannot be altered \textit{in situ}.
Given a suitable gate pattern, the interaction matrix elements $V_{ij}$, the local energy offsets $\varepsilon_i$ and nearest-neighbour hopping $t_{ij}$ may in principle still be tuned independently from one another~\cite{Scarlino2021}.
For simplicity, we fix $t$ and $V_0$ in our numerical simulation and treat them as constant for all distances $R$.
Despite the fact that, in this way, we do not optimize the scaling parameter $\eta$ for each $R$ separately, we obtain a molecular binding curve with a pronounced minimum as shown in Fig.~\ref{fig:chemistry}(d).
We show the molecular binding energy $\Delta = E_{2\mathrm{e}^-} - 2E_{\mathrm{e}^-}$, and thus compare the ground-state energies of the two-electron ($E_{2\mathrm{e}^-}$) and single-electron ($E_{\mathrm{e}^-}$) systems.
The results show that basic molecular properties can be resolved with arrays of moderate size, e.g., $N \approx 10$ dots suffice for determining the bond length of the artificial molecule.

\subsection{Relevant experimental techniques \label{ssec:experimental-techniques}}
For the experimental implementation it is important to consider methods for the initialization of the QC simulator and the measurement of relevant system parameters.
The initialization requires the QD array to be occupied with a fixed total number of charges.
For small systems this can be achieved based on that the long-range electron-electron interaction strength is larger than the reservoir temperature. 
For a larger number of sites with only few charges, control over the tunnel coupling to the reservoirs can allow for first loading the desired number of charges and then isolating the QD array by raising the respective tunnel barriers~\cite{Bertrand2015, Bayer2019}.

For measurements of the simulator there are various techniques available for gate-defined QDs.
The energy splittings for the artificial atom and molecule, see Fig.~\ref{fig:chemistry}, are found to be in the \SIrange{5}{60}{\micro \electronvolt} range, and thus can be probed with microwave spectroscopy such as used for the observation of covalent bonding on a DQD~\cite{Oosterkamp1998}. 
The microwave excitation and charge sensor detection require a change in charge distribution between ground and excited state, which are shown in Fig.~\ref{fig:schematic}(c) and (d) for the atom and molecular ion, respectively.
The transition probability can be increased by applying the microwave signal on multiple gates. 
In addition, a tilt on local offsets $\varepsilon_i$ could be applied, see App.~\ref{app:occ_num_mol}, which has a similar effect as an electric field that induces a dc Stark shift for an atom or molecule.
This could in particular prove useful for the simulation of molecules with larger internuclear separation, see Fig.~\ref{fig:app_occupation}.
Measurements of the energy levels of the QC simulator, for example for the ground state of the two-electron molecule shown in Fig.~\ref{fig:chemistry}(d), could be performed by using a well-defined reference level, e.g., the reservoir Fermi level or another QD. The energy level of the QC simulator can then be extracted by identifying for which global shift the simulator level is on resonance with the reference level. 
Another useful technique for QDs is gate-based reflectometry by which the electrical susceptibility can be measured~\cite{Pakkiam2018, Urdampilleta2019, West2019, Zheng2019}.

\section{Conclusions \& outlook \label{sec:outlook}}
In summary, we have experimentally characterized the long-range electron-electron interactions in a gate-defined quantum dot array.
The interactions were experimentally found to be detectable between electrons up to four sites away.
We compared a toy model of the electrostatic interaction that considers metallic gates as the main screening source and found good agreement with our measurement results.
In future work it will be instructive to analyse in more detail how the form of long-range interaction potentials can be controlled by an adequate choice of gate patterns.

We have also discussed how quantum-dot systems may be utilized for future analog simulations of artificial quantum matter, both in single-electron systems and in many-electron systems with long-range interactions.
Using numerical simulations, we demonstrated that quantized atomic binding energies can be resolved and that the interactions are sufficiently strong to explore non-trivial properties like molecular dissociation curves.
We have shown that proof-of-principle demonstrations may already be performed using state-of-the-art experimental setups of $\approx 10$ quantum dots.
Hence our study opens up the path for future quantum simulation experiments and studies of artificial atoms and molecules using semiconductor quantum dot arrays.
Due to the computationally challenging problems that arise in the context of quantum chemistry, this may prove beneficial for the benchmarking of existing numerical techniques and development of new theoretical methods.
In addition, these well-controlled quantum systems with long-range interactions are also promising for future investigations of other debated phenomena, such as Wigner crystallization~\cite{Vu2020, Vu2020b}, exciton formation~\cite{Frenkel1931}, high-temperature superconductivity~\cite{Manousakis2002, Anderson2013}, and the nature of many-body excited states \cite{Townsend2021}.

The data reported in this paper and scripts to generate the figures are archived at doi.org/10.5281/zenodo.6035944.

\section*{Acknowledgements}
J.K. and J.I.C. acknowledge support from the Deutsche Forschungsgemeinschaft (DFG, German Research Foundation) under Germany’s Excellence Strategy – EXC-2111 – 390814868, and from ERC Advanced Grant QUENOCOBA under the EU Horizon 2020 program (Grant Agreement No. 742102).
G.G. acknowledges support by the Spanish AEI through project
no.~PID2020-115406GB-I00 ``{GEQCO}'' and the European FET-OPEN project SPRING (No. 863098).
L.M.K.V. acknowledges support from the European Research Council (ERC Advanced Grant, grant number 882848 — QuDoFH). C.R. and W.W. acknowledge support from the Swiss National Science Foundation.

\bibliographystyle{apsrev}
\bibliography{main}

\begin{thebibliography}{43}
\expandafter\ifx\csname natexlab\endcsname\relax\def\natexlab#1{#1}\fi
\expandafter\ifx\csname bibnamefont\endcsname\relax
  \def\bibnamefont#1{#1}\fi
\expandafter\ifx\csname bibfnamefont\endcsname\relax
  \def\bibfnamefont#1{#1}\fi
\expandafter\ifx\csname citenamefont\endcsname\relax
  \def\citenamefont#1{#1}\fi
\expandafter\ifx\csname url\endcsname\relax
  \def\url#1{\texttt{#1}}\fi
\expandafter\ifx\csname urlprefix\endcsname\relax\def\urlprefix{URL }\fi
\providecommand{\bibinfo}[2]{#2}
\providecommand{\eprint}[2][]{\url{#2}}

\bibitem[{\citenamefont{French et~al.}(2010)\citenamefont{French, Parsegian,
  Podgornik, Rajter, Jagota, Luo, Asthagiri, Chaudhury, Chiang, Granick
  et~al.}}]{French2010}
\bibinfo{author}{\bibfnamefont{R.~H.} \bibnamefont{French}},
  \bibinfo{author}{\bibfnamefont{V.~A.} \bibnamefont{Parsegian}},
  \bibinfo{author}{\bibfnamefont{R.}~\bibnamefont{Podgornik}},
  \bibinfo{author}{\bibfnamefont{R.~F.} \bibnamefont{Rajter}},
  \bibinfo{author}{\bibfnamefont{A.}~\bibnamefont{Jagota}},
  \bibinfo{author}{\bibfnamefont{J.}~\bibnamefont{Luo}},
  \bibinfo{author}{\bibfnamefont{D.}~\bibnamefont{Asthagiri}},
  \bibinfo{author}{\bibfnamefont{M.~K.} \bibnamefont{Chaudhury}},
  \bibinfo{author}{\bibfnamefont{Y.-m.} \bibnamefont{Chiang}},
  \bibinfo{author}{\bibfnamefont{S.}~\bibnamefont{Granick}},
  \bibnamefont{et~al.}, \bibinfo{journal}{Rev. Mod. Phys.}
  \textbf{\bibinfo{volume}{82}}, \bibinfo{pages}{1887} (\bibinfo{year}{2010}).

\bibitem[{\citenamefont{Wigner}(1934)}]{Wigner1934}
\bibinfo{author}{\bibfnamefont{E.}~\bibnamefont{Wigner}},
  \bibinfo{journal}{Phys. Rev.} \textbf{\bibinfo{volume}{46}},
  \bibinfo{pages}{1002} (\bibinfo{year}{1934}).

\bibitem[{\citenamefont{Frenkel}(1931)}]{Frenkel1931}
\bibinfo{author}{\bibfnamefont{J.}~\bibnamefont{Frenkel}},
  \bibinfo{journal}{Phys. Rev.} \textbf{\bibinfo{volume}{37}},
  \bibinfo{pages}{17} (\bibinfo{year}{1931}).

\bibitem[{\citenamefont{Anderson}(2013)}]{Anderson2013}
\bibinfo{author}{\bibfnamefont{P.~W.} \bibnamefont{Anderson}},
  \bibinfo{journal}{J. Phys.: Conf. Ser.} \textbf{\bibinfo{volume}{449}},
  \bibinfo{pages}{012001} (\bibinfo{year}{2013}).

\bibitem[{\citenamefont{Cirac and Zoller}(2012)}]{Cirac2012}
\bibinfo{author}{\bibfnamefont{J.~I.} \bibnamefont{Cirac}} \bibnamefont{and}
  \bibinfo{author}{\bibfnamefont{P.}~\bibnamefont{Zoller}},
  \bibinfo{journal}{Nat. Phys.} \textbf{\bibinfo{volume}{8}},
  \bibinfo{pages}{264} (\bibinfo{year}{2012}).

\bibitem[{\citenamefont{Georgescu et~al.}(2014)\citenamefont{Georgescu, Ashhab,
  and Nori}}]{Georgescu2014}
\bibinfo{author}{\bibfnamefont{I.~M.} \bibnamefont{Georgescu}},
  \bibinfo{author}{\bibfnamefont{S.}~\bibnamefont{Ashhab}}, \bibnamefont{and}
  \bibinfo{author}{\bibfnamefont{F.}~\bibnamefont{Nori}},
  \bibinfo{journal}{Rev. Mod. Phys.} \textbf{\bibinfo{volume}{86}},
  \bibinfo{pages}{153} (\bibinfo{year}{2014}).

\bibitem[{\citenamefont{Manousakis}(2002)}]{Manousakis2002}
\bibinfo{author}{\bibfnamefont{E.}~\bibnamefont{Manousakis}},
  \bibinfo{journal}{J. Low Temp. Phys.} \textbf{\bibinfo{volume}{126}},
  \bibinfo{pages}{1501} (\bibinfo{year}{2002}).

\bibitem[{\citenamefont{Byrnes et~al.}(2008)\citenamefont{Byrnes, Kim, Kusudo,
  and Yamamoto}}]{Byrnes2008}
\bibinfo{author}{\bibfnamefont{T.}~\bibnamefont{Byrnes}},
  \bibinfo{author}{\bibfnamefont{N.~Y.} \bibnamefont{Kim}},
  \bibinfo{author}{\bibfnamefont{K.}~\bibnamefont{Kusudo}}, \bibnamefont{and}
  \bibinfo{author}{\bibfnamefont{Y.}~\bibnamefont{Yamamoto}},
  \bibinfo{journal}{Phys. Rev. B} \textbf{\bibinfo{volume}{78}},
  \bibinfo{pages}{075320} (\bibinfo{year}{2008}).

\bibitem[{\citenamefont{Hensgens et~al.}(2017)\citenamefont{Hensgens, Fujita,
  Janssen, Li, {Van Diepen}, Reichl, Wegscheider, {Das Sarma}, and
  Vandersypen}}]{Hensgens2017}
\bibinfo{author}{\bibfnamefont{T.}~\bibnamefont{Hensgens}},
  \bibinfo{author}{\bibfnamefont{T.}~\bibnamefont{Fujita}},
  \bibinfo{author}{\bibfnamefont{L.}~\bibnamefont{Janssen}},
  \bibinfo{author}{\bibfnamefont{X.}~\bibnamefont{Li}},
  \bibinfo{author}{\bibfnamefont{C.~J.} \bibnamefont{{Van Diepen}}},
  \bibinfo{author}{\bibfnamefont{C.}~\bibnamefont{Reichl}},
  \bibinfo{author}{\bibfnamefont{W.}~\bibnamefont{Wegscheider}},
  \bibinfo{author}{\bibfnamefont{S.}~\bibnamefont{{Das Sarma}}},
  \bibnamefont{and} \bibinfo{author}{\bibfnamefont{L.~M.~K.}
  \bibnamefont{Vandersypen}}, \bibinfo{journal}{Nature}
  \textbf{\bibinfo{volume}{548}}, \bibinfo{pages}{70} (\bibinfo{year}{2017}).

\bibitem[{\citenamefont{Dehollain et~al.}(2020)\citenamefont{Dehollain,
  Mukhopadhyay, Michal, Wang, Wunsch, Reichl, Wegscheider, Rudner, Demler, and
  Vandersypen}}]{Dehollain2020}
\bibinfo{author}{\bibfnamefont{J.~P.} \bibnamefont{Dehollain}},
  \bibinfo{author}{\bibfnamefont{U.}~\bibnamefont{Mukhopadhyay}},
  \bibinfo{author}{\bibfnamefont{V.~P.} \bibnamefont{Michal}},
  \bibinfo{author}{\bibfnamefont{Y.}~\bibnamefont{Wang}},
  \bibinfo{author}{\bibfnamefont{B.}~\bibnamefont{Wunsch}},
  \bibinfo{author}{\bibfnamefont{C.}~\bibnamefont{Reichl}},
  \bibinfo{author}{\bibfnamefont{W.}~\bibnamefont{Wegscheider}},
  \bibinfo{author}{\bibfnamefont{M.~S.} \bibnamefont{Rudner}},
  \bibinfo{author}{\bibfnamefont{E.}~\bibnamefont{Demler}}, \bibnamefont{and}
  \bibinfo{author}{\bibfnamefont{L.~M.~K.} \bibnamefont{Vandersypen}},
  \bibinfo{journal}{Nature} \textbf{\bibinfo{volume}{579}},
  \bibinfo{pages}{528} (\bibinfo{year}{2020}).

\bibitem[{\citenamefont{van Diepen et~al.}(2021{\natexlab{a}})\citenamefont{van
  Diepen, Hsiao, Mukhopadhyay, Reichl, Wegscheider, and
  Vandersypen}}]{VanDiepen2021b}
\bibinfo{author}{\bibfnamefont{C.~J.} \bibnamefont{van Diepen}},
  \bibinfo{author}{\bibfnamefont{T.-K.} \bibnamefont{Hsiao}},
  \bibinfo{author}{\bibfnamefont{U.}~\bibnamefont{Mukhopadhyay}},
  \bibinfo{author}{\bibfnamefont{C.}~\bibnamefont{Reichl}},
  \bibinfo{author}{\bibfnamefont{W.}~\bibnamefont{Wegscheider}},
  \bibnamefont{and} \bibinfo{author}{\bibfnamefont{L.~M.~K.}
  \bibnamefont{Vandersypen}}, \bibinfo{journal}{Phys. Rev. X}
  \textbf{\bibinfo{volume}{11}}, \bibinfo{pages}{041025}
  (\bibinfo{year}{2021}{\natexlab{a}}).

\bibitem[{\citenamefont{Li et~al.}(2018)\citenamefont{Li, Petit, Franke,
  Dehollain, Helsen, Steudtner, Thomas, Yoscovits, Singh, Wehner
  et~al.}}]{Li2018}
\bibinfo{author}{\bibfnamefont{R.}~\bibnamefont{Li}},
  \bibinfo{author}{\bibfnamefont{L.}~\bibnamefont{Petit}},
  \bibinfo{author}{\bibfnamefont{D.~P.} \bibnamefont{Franke}},
  \bibinfo{author}{\bibfnamefont{J.~P.} \bibnamefont{Dehollain}},
  \bibinfo{author}{\bibfnamefont{J.}~\bibnamefont{Helsen}},
  \bibinfo{author}{\bibfnamefont{M.}~\bibnamefont{Steudtner}},
  \bibinfo{author}{\bibfnamefont{N.~K.} \bibnamefont{Thomas}},
  \bibinfo{author}{\bibfnamefont{Z.~R.} \bibnamefont{Yoscovits}},
  \bibinfo{author}{\bibfnamefont{K.~J.} \bibnamefont{Singh}},
  \bibinfo{author}{\bibfnamefont{S.}~\bibnamefont{Wehner}},
  \bibnamefont{et~al.}, \bibinfo{journal}{Sci. Adv.}
  \textbf{\bibinfo{volume}{4}}, \bibinfo{pages}{eaar3960}
  (\bibinfo{year}{2018}).

\bibitem[{\citenamefont{van~der Wiel et~al.}(2003)\citenamefont{van~der Wiel,
  De~Franceschi, Elzerman, Fujisawa, Tarucha, and Kouwenhoven}}]{Wiel2003}
\bibinfo{author}{\bibfnamefont{W.~G.} \bibnamefont{van~der Wiel}},
  \bibinfo{author}{\bibfnamefont{S.}~\bibnamefont{De~Franceschi}},
  \bibinfo{author}{\bibfnamefont{J.~M.} \bibnamefont{Elzerman}},
  \bibinfo{author}{\bibfnamefont{T.}~\bibnamefont{Fujisawa}},
  \bibinfo{author}{\bibfnamefont{S.}~\bibnamefont{Tarucha}}, \bibnamefont{and}
  \bibinfo{author}{\bibfnamefont{L.~P.} \bibnamefont{Kouwenhoven}},
  \bibinfo{journal}{Rev. Mod. Phys.} \textbf{\bibinfo{volume}{75}},
  \bibinfo{pages}{1} (\bibinfo{year}{2003}).

\bibitem[{\citenamefont{Shulman et~al.}(2012)\citenamefont{Shulman, Dial,
  Harvey, Bluhm, Umansky, and Yacoby}}]{Shulman2012}
\bibinfo{author}{\bibfnamefont{M.~D.} \bibnamefont{Shulman}},
  \bibinfo{author}{\bibfnamefont{O.~E.} \bibnamefont{Dial}},
  \bibinfo{author}{\bibfnamefont{S.~P.} \bibnamefont{Harvey}},
  \bibinfo{author}{\bibfnamefont{H.}~\bibnamefont{Bluhm}},
  \bibinfo{author}{\bibfnamefont{V.}~\bibnamefont{Umansky}}, \bibnamefont{and}
  \bibinfo{author}{\bibfnamefont{A.}~\bibnamefont{Yacoby}},
  \bibinfo{journal}{Science} \textbf{\bibinfo{volume}{336}},
  \bibinfo{pages}{202} (\bibinfo{year}{2012}).

\bibitem[{\citenamefont{Zajac et~al.}(2016)\citenamefont{Zajac, Hazard, Mi,
  Nielsen, and Petta}}]{Zajac2016}
\bibinfo{author}{\bibfnamefont{D.~M.} \bibnamefont{Zajac}},
  \bibinfo{author}{\bibfnamefont{T.~M.} \bibnamefont{Hazard}},
  \bibinfo{author}{\bibfnamefont{X.}~\bibnamefont{Mi}},
  \bibinfo{author}{\bibfnamefont{E.}~\bibnamefont{Nielsen}}, \bibnamefont{and}
  \bibinfo{author}{\bibfnamefont{J.~R.} \bibnamefont{Petta}},
  \bibinfo{journal}{Phys. Rev. Appl.} \textbf{\bibinfo{volume}{6}},
  \bibinfo{pages}{054013} (\bibinfo{year}{2016}).

\bibitem[{\citenamefont{Neyens et~al.}(2019)\citenamefont{Neyens, MacQuarrie,
  Dodson, Corrigan, Holman, Thorgrimsson, Palma, McJunkin, Edge, Friesen
  et~al.}}]{Neyens2019}
\bibinfo{author}{\bibfnamefont{S.~F.} \bibnamefont{Neyens}},
  \bibinfo{author}{\bibfnamefont{E.~R.} \bibnamefont{MacQuarrie}},
  \bibinfo{author}{\bibfnamefont{J.~P.} \bibnamefont{Dodson}},
  \bibinfo{author}{\bibfnamefont{J.}~\bibnamefont{Corrigan}},
  \bibinfo{author}{\bibfnamefont{N.}~\bibnamefont{Holman}},
  \bibinfo{author}{\bibfnamefont{B.}~\bibnamefont{Thorgrimsson}},
  \bibinfo{author}{\bibfnamefont{M.}~\bibnamefont{Palma}},
  \bibinfo{author}{\bibfnamefont{T.}~\bibnamefont{McJunkin}},
  \bibinfo{author}{\bibfnamefont{L.~F.} \bibnamefont{Edge}},
  \bibinfo{author}{\bibfnamefont{M.}~\bibnamefont{Friesen}},
  \bibnamefont{et~al.}, \bibinfo{journal}{Phys. Rev. Appl.}
  \textbf{\bibinfo{volume}{12}}, \bibinfo{pages}{064049}
  (\bibinfo{year}{2019}).

\bibitem[{\citenamefont{van Diepen et~al.}(2021{\natexlab{b}})\citenamefont{van
  Diepen, Hsiao, Mukhopadhyay, Reichl, Wegscheider, and
  Vandersypen}}]{VanDiepen2021}
\bibinfo{author}{\bibfnamefont{C.~J.} \bibnamefont{van Diepen}},
  \bibinfo{author}{\bibfnamefont{T.-K.} \bibnamefont{Hsiao}},
  \bibinfo{author}{\bibfnamefont{U.}~\bibnamefont{Mukhopadhyay}},
  \bibinfo{author}{\bibfnamefont{C.}~\bibnamefont{Reichl}},
  \bibinfo{author}{\bibfnamefont{W.}~\bibnamefont{Wegscheider}},
  \bibnamefont{and} \bibinfo{author}{\bibfnamefont{L.~M.~K.}
  \bibnamefont{Vandersypen}}, \bibinfo{journal}{Nat. Commun.}
  \textbf{\bibinfo{volume}{12}}, \bibinfo{pages}{77}
  (\bibinfo{year}{2021}{\natexlab{b}}).

\bibitem[{\citenamefont{Volk et~al.}(2019)\citenamefont{Volk, Zwerver,
  Mukhopadhyay, Eendebak, van Diepen, Dehollain, Hensgens, Fujita, Reichl,
  Wegscheider et~al.}}]{Volk2019}
\bibinfo{author}{\bibfnamefont{C.}~\bibnamefont{Volk}},
  \bibinfo{author}{\bibfnamefont{A.~M.~J.} \bibnamefont{Zwerver}},
  \bibinfo{author}{\bibfnamefont{U.}~\bibnamefont{Mukhopadhyay}},
  \bibinfo{author}{\bibfnamefont{P.~T.} \bibnamefont{Eendebak}},
  \bibinfo{author}{\bibfnamefont{C.~J.} \bibnamefont{van Diepen}},
  \bibinfo{author}{\bibfnamefont{J.~P.} \bibnamefont{Dehollain}},
  \bibinfo{author}{\bibfnamefont{T.}~\bibnamefont{Hensgens}},
  \bibinfo{author}{\bibfnamefont{T.}~\bibnamefont{Fujita}},
  \bibinfo{author}{\bibfnamefont{C.}~\bibnamefont{Reichl}},
  \bibinfo{author}{\bibfnamefont{W.}~\bibnamefont{Wegscheider}},
  \bibnamefont{et~al.}, \bibinfo{journal}{npj Quantum Inf.}
  \textbf{\bibinfo{volume}{5}}, \bibinfo{pages}{29} (\bibinfo{year}{2019}).

\bibitem[{\citenamefont{Mills et~al.}(2019)\citenamefont{Mills, Zajac, Gullans,
  Schupp, Hazard, and Petta}}]{Mills2019}
\bibinfo{author}{\bibfnamefont{A.~R.} \bibnamefont{Mills}},
  \bibinfo{author}{\bibfnamefont{D.~M.} \bibnamefont{Zajac}},
  \bibinfo{author}{\bibfnamefont{M.~J.} \bibnamefont{Gullans}},
  \bibinfo{author}{\bibfnamefont{F.~J.} \bibnamefont{Schupp}},
  \bibinfo{author}{\bibfnamefont{T.~M.} \bibnamefont{Hazard}},
  \bibnamefont{and} \bibinfo{author}{\bibfnamefont{J.~R.} \bibnamefont{Petta}},
  \bibinfo{journal}{Nat. Commun.} \textbf{\bibinfo{volume}{10}}
  (\bibinfo{year}{2019}).

\bibitem[{\citenamefont{Hsiao et~al.}(2020)\citenamefont{Hsiao, van Diepen,
  Mukhopadhyay, Reichl, Wegscheider, and Vandersypen}}]{Hsiao2020}
\bibinfo{author}{\bibfnamefont{T.-K.} \bibnamefont{Hsiao}},
  \bibinfo{author}{\bibfnamefont{C.~J.} \bibnamefont{van Diepen}},
  \bibinfo{author}{\bibfnamefont{U.}~\bibnamefont{Mukhopadhyay}},
  \bibinfo{author}{\bibfnamefont{C.}~\bibnamefont{Reichl}},
  \bibinfo{author}{\bibfnamefont{W.}~\bibnamefont{Wegscheider}},
  \bibnamefont{and} \bibinfo{author}{\bibfnamefont{L.~M.~K.}
  \bibnamefont{Vandersypen}}, \bibinfo{journal}{Phys. Rev. Appl.}
  \textbf{\bibinfo{volume}{13}}, \bibinfo{pages}{054018}
  (\bibinfo{year}{2020}).

\bibitem[{\citenamefont{Qiao et~al.}(2020)\citenamefont{Qiao, Kandel, Deng,
  Fallahi, Gardner, Manfra, Barnes, and Nichol}}]{Qiao2020}
\bibinfo{author}{\bibfnamefont{H.}~\bibnamefont{Qiao}},
  \bibinfo{author}{\bibfnamefont{Y.~P.} \bibnamefont{Kandel}},
  \bibinfo{author}{\bibfnamefont{K.}~\bibnamefont{Deng}},
  \bibinfo{author}{\bibfnamefont{S.}~\bibnamefont{Fallahi}},
  \bibinfo{author}{\bibfnamefont{G.~C.} \bibnamefont{Gardner}},
  \bibinfo{author}{\bibfnamefont{M.~J.} \bibnamefont{Manfra}},
  \bibinfo{author}{\bibfnamefont{E.}~\bibnamefont{Barnes}}, \bibnamefont{and}
  \bibinfo{author}{\bibfnamefont{J.~M.} \bibnamefont{Nichol}},
  \bibinfo{journal}{Phys. Rev. X} \textbf{\bibinfo{volume}{10}},
  \bibinfo{pages}{31006} (\bibinfo{year}{2020}), \eprint{2001.02277}.

\bibitem[{\citenamefont{Arg{\"{u}}ello-Luengo
  et~al.}(2019)\citenamefont{Arg{\"{u}}ello-Luengo, Gonz{\'{a}}lez-Tudela, Shi,
  Zoller, and Cirac}}]{Arguello2019}
\bibinfo{author}{\bibfnamefont{J.}~\bibnamefont{Arg{\"{u}}ello-Luengo}},
  \bibinfo{author}{\bibfnamefont{A.}~\bibnamefont{Gonz{\'{a}}lez-Tudela}},
  \bibinfo{author}{\bibfnamefont{T.}~\bibnamefont{Shi}},
  \bibinfo{author}{\bibfnamefont{P.}~\bibnamefont{Zoller}}, \bibnamefont{and}
  \bibinfo{author}{\bibfnamefont{J.~I.} \bibnamefont{Cirac}},
  \bibinfo{journal}{Nature} \textbf{\bibinfo{volume}{574}},
  \bibinfo{pages}{215} (\bibinfo{year}{2019}).

\bibitem[{\citenamefont{Loudon}(2016)}]{Loudon2016}
\bibinfo{author}{\bibfnamefont{R.}~\bibnamefont{Loudon}},
  \bibinfo{journal}{Proc. R. Soc. A} \textbf{\bibinfo{volume}{472}},
  \bibinfo{pages}{2050534} (\bibinfo{year}{2016}).

\bibitem[{\citenamefont{Loos et~al.}(2015)\citenamefont{Loos, Ball, and
  Gill}}]{Loos2015}
\bibinfo{author}{\bibfnamefont{P.-F.} \bibnamefont{Loos}},
  \bibinfo{author}{\bibfnamefont{C.~J.} \bibnamefont{Ball}}, \bibnamefont{and}
  \bibinfo{author}{\bibfnamefont{P.~M.~W.} \bibnamefont{Gill}},
  \bibinfo{journal}{Phys. Chem. Chem. Phys.} \textbf{\bibinfo{volume}{17}},
  \bibinfo{pages}{3196} (\bibinfo{year}{2015}).

\bibitem[{\citenamefont{Imada et~al.}(1998)\citenamefont{Imada, Fujimori, and
  Tokura}}]{Imada1998}
\bibinfo{author}{\bibfnamefont{M.}~\bibnamefont{Imada}},
  \bibinfo{author}{\bibfnamefont{A.}~\bibnamefont{Fujimori}}, \bibnamefont{and}
  \bibinfo{author}{\bibfnamefont{Y.}~\bibnamefont{Tokura}},
  \bibinfo{journal}{Rev. Mod. Phys.} \textbf{\bibinfo{volume}{70}},
  \bibinfo{pages}{1039} (\bibinfo{year}{1998}).

\bibitem[{\citenamefont{Yang et~al.}(2011)\citenamefont{Yang, Wang, and {Das
  Sarma}}}]{Yang2011}
\bibinfo{author}{\bibfnamefont{S.}~\bibnamefont{Yang}},
  \bibinfo{author}{\bibfnamefont{X.}~\bibnamefont{Wang}}, \bibnamefont{and}
  \bibinfo{author}{\bibfnamefont{S.}~\bibnamefont{{Das Sarma}}},
  \bibinfo{journal}{Phys. Rev. B} \textbf{\bibinfo{volume}{83}},
  \bibinfo{pages}{161301} (\bibinfo{year}{2011}).

\bibitem[{\citenamefont{{Van Diepen} et~al.}(2018)\citenamefont{{Van Diepen},
  Eendebak, Buijtendorp, Mukhopadhyay, Fujita, Reichl, Wegscheider, and
  Vandersypen}}]{VanDiepen2018}
\bibinfo{author}{\bibfnamefont{C.~J.} \bibnamefont{{Van Diepen}}},
  \bibinfo{author}{\bibfnamefont{P.~T.} \bibnamefont{Eendebak}},
  \bibinfo{author}{\bibfnamefont{B.~T.} \bibnamefont{Buijtendorp}},
  \bibinfo{author}{\bibfnamefont{U.}~\bibnamefont{Mukhopadhyay}},
  \bibinfo{author}{\bibfnamefont{T.}~\bibnamefont{Fujita}},
  \bibinfo{author}{\bibfnamefont{C.}~\bibnamefont{Reichl}},
  \bibinfo{author}{\bibfnamefont{W.}~\bibnamefont{Wegscheider}},
  \bibnamefont{and} \bibinfo{author}{\bibfnamefont{L.~M.~K.}
  \bibnamefont{Vandersypen}}, \bibinfo{journal}{Appl. Phys. Lett.}
  \textbf{\bibinfo{volume}{113}}, \bibinfo{pages}{033101}
  (\bibinfo{year}{2018}).

\bibitem[{\citenamefont{Nowack et~al.}(2011)\citenamefont{Nowack, Shafiei,
  Laforest, Prawiroatmodjo, Schreiber, Reichl, Wegscheider, and
  Vandersypen}}]{Nowack2011}
\bibinfo{author}{\bibfnamefont{K.~C.} \bibnamefont{Nowack}},
  \bibinfo{author}{\bibfnamefont{M.}~\bibnamefont{Shafiei}},
  \bibinfo{author}{\bibfnamefont{M.}~\bibnamefont{Laforest}},
  \bibinfo{author}{\bibfnamefont{G.~E. D.~K.} \bibnamefont{Prawiroatmodjo}},
  \bibinfo{author}{\bibfnamefont{L.~R.} \bibnamefont{Schreiber}},
  \bibinfo{author}{\bibfnamefont{C.}~\bibnamefont{Reichl}},
  \bibinfo{author}{\bibfnamefont{W.}~\bibnamefont{Wegscheider}},
  \bibnamefont{and} \bibinfo{author}{\bibfnamefont{L.~M.~K.}
  \bibnamefont{Vandersypen}}, \bibinfo{journal}{Science}
  \textbf{\bibinfo{volume}{333}}, \bibinfo{pages}{1269} (\bibinfo{year}{2011}).

\bibitem[{\citenamefont{OpenCV}(2015)}]{2015opencv}
\bibinfo{author}{\bibnamefont{OpenCV}}, \emph{\bibinfo{title}{Open source
  computer vision library}} (\bibinfo{year}{2015}).

\bibitem[{\citenamefont{Segal et~al.}(2004)\citenamefont{Segal, Kr\'al, and
  Shapiro}}]{Segal2004}
\bibinfo{author}{\bibfnamefont{D.}~\bibnamefont{Segal}},
  \bibinfo{author}{\bibfnamefont{P.}~\bibnamefont{Kr\'al}}, \bibnamefont{and}
  \bibinfo{author}{\bibfnamefont{M.}~\bibnamefont{Shapiro}},
  \bibinfo{journal}{Phys. Rev. B} \textbf{\bibinfo{volume}{69}},
  \bibinfo{pages}{153405} (\bibinfo{year}{2004}).

\bibitem[{\citenamefont{Lawrie et~al.}(2020)\citenamefont{Lawrie, Eenink,
  Hendrickx, Boter, Petit, Amitonov, Lodari, {Paquelet Wuetz}, Volk, Philips
  et~al.}}]{Lawrie2020}
\bibinfo{author}{\bibfnamefont{W.~I.~L.} \bibnamefont{Lawrie}},
  \bibinfo{author}{\bibfnamefont{H.~G.~J.} \bibnamefont{Eenink}},
  \bibinfo{author}{\bibfnamefont{N.~W.} \bibnamefont{Hendrickx}},
  \bibinfo{author}{\bibfnamefont{J.~M.} \bibnamefont{Boter}},
  \bibinfo{author}{\bibfnamefont{L.}~\bibnamefont{Petit}},
  \bibinfo{author}{\bibfnamefont{S.~V.} \bibnamefont{Amitonov}},
  \bibinfo{author}{\bibfnamefont{M.}~\bibnamefont{Lodari}},
  \bibinfo{author}{\bibfnamefont{B.}~\bibnamefont{{Paquelet Wuetz}}},
  \bibinfo{author}{\bibfnamefont{C.}~\bibnamefont{Volk}},
  \bibinfo{author}{\bibfnamefont{S.~G.~J.} \bibnamefont{Philips}},
  \bibnamefont{et~al.}, \bibinfo{journal}{Appl. Phys. Lett.}
  \textbf{\bibinfo{volume}{116}} (\bibinfo{year}{2020}).

\bibitem[{\citenamefont{Arg{\"u}ello-Luengo
  et~al.}(2020)\citenamefont{Arg{\"u}ello-Luengo, Gonz{\'a}lez-Tudela, Shi,
  Zoller, and Cirac}}]{Arguello2020}
\bibinfo{author}{\bibfnamefont{J.}~\bibnamefont{Arg{\"u}ello-Luengo}},
  \bibinfo{author}{\bibfnamefont{A.}~\bibnamefont{Gonz{\'a}lez-Tudela}},
  \bibinfo{author}{\bibfnamefont{T.}~\bibnamefont{Shi}},
  \bibinfo{author}{\bibfnamefont{P.}~\bibnamefont{Zoller}}, \bibnamefont{and}
  \bibinfo{author}{\bibfnamefont{J.~I.} \bibnamefont{Cirac}},
  \bibinfo{journal}{Phys. Rev. Res.} \textbf{\bibinfo{volume}{2}},
  \bibinfo{pages}{042013} (\bibinfo{year}{2020}).

\bibitem[{\citenamefont{Scarlino et~al.}(2021)\citenamefont{Scarlino, Ungerer,
  van Woerkom, Mancini, Stano, Muller, Landig, Koski, Reichl, Wegscheider
  et~al.}}]{Scarlino2021}
\bibinfo{author}{\bibfnamefont{P.}~\bibnamefont{Scarlino}},
  \bibinfo{author}{\bibfnamefont{J.~H.} \bibnamefont{Ungerer}},
  \bibinfo{author}{\bibfnamefont{D.~J.} \bibnamefont{van Woerkom}},
  \bibinfo{author}{\bibfnamefont{M.}~\bibnamefont{Mancini}},
  \bibinfo{author}{\bibfnamefont{P.}~\bibnamefont{Stano}},
  \bibinfo{author}{\bibfnamefont{C.}~\bibnamefont{Muller}},
  \bibinfo{author}{\bibfnamefont{A.~J.} \bibnamefont{Landig}},
  \bibinfo{author}{\bibfnamefont{J.~V.} \bibnamefont{Koski}},
  \bibinfo{author}{\bibfnamefont{C.}~\bibnamefont{Reichl}},
  \bibinfo{author}{\bibfnamefont{W.}~\bibnamefont{Wegscheider}},
  \bibnamefont{et~al.}, \bibinfo{journal}{arXiv:2104.03045}
  (\bibinfo{year}{2021}).

\bibitem[{\citenamefont{Bertrand et~al.}(2015)\citenamefont{Bertrand, Flentje,
  Takada, Yamamoto, Tarucha, Ludwig, Wieck, B\"auerle, and
  Meunier}}]{Bertrand2015}
\bibinfo{author}{\bibfnamefont{B.}~\bibnamefont{Bertrand}},
  \bibinfo{author}{\bibfnamefont{H.}~\bibnamefont{Flentje}},
  \bibinfo{author}{\bibfnamefont{S.}~\bibnamefont{Takada}},
  \bibinfo{author}{\bibfnamefont{M.}~\bibnamefont{Yamamoto}},
  \bibinfo{author}{\bibfnamefont{S.}~\bibnamefont{Tarucha}},
  \bibinfo{author}{\bibfnamefont{A.}~\bibnamefont{Ludwig}},
  \bibinfo{author}{\bibfnamefont{A.~D.} \bibnamefont{Wieck}},
  \bibinfo{author}{\bibfnamefont{C.}~\bibnamefont{B\"auerle}},
  \bibnamefont{and} \bibinfo{author}{\bibfnamefont{T.}~\bibnamefont{Meunier}},
  \bibinfo{journal}{Phys. Rev. Lett.} \textbf{\bibinfo{volume}{115}},
  \bibinfo{pages}{1} (\bibinfo{year}{2015}).

\bibitem[{\citenamefont{Bayer et~al.}(2019)\citenamefont{Bayer, Wagner,
  Rugeramigabo, and Haug}}]{Bayer2019}
\bibinfo{author}{\bibfnamefont{J.~C.} \bibnamefont{Bayer}},
  \bibinfo{author}{\bibfnamefont{T.}~\bibnamefont{Wagner}},
  \bibinfo{author}{\bibfnamefont{E.~P.} \bibnamefont{Rugeramigabo}},
  \bibnamefont{and} \bibinfo{author}{\bibfnamefont{R.~J.} \bibnamefont{Haug}},
  \bibinfo{journal}{Ann. Phys.} \textbf{\bibinfo{volume}{531}},
  \bibinfo{pages}{1} (\bibinfo{year}{2019}).

\bibitem[{\citenamefont{Oosterkamp et~al.}(1998)\citenamefont{Oosterkamp,
  Fujisawa, Wiel, Ishibashi, Hijman, Tarucha, and
  Kouwenhoven}}]{Oosterkamp1998}
\bibinfo{author}{\bibfnamefont{T.~H.} \bibnamefont{Oosterkamp}},
  \bibinfo{author}{\bibfnamefont{T.}~\bibnamefont{Fujisawa}},
  \bibinfo{author}{\bibfnamefont{W.~G. V.~D.} \bibnamefont{Wiel}},
  \bibinfo{author}{\bibfnamefont{K.}~\bibnamefont{Ishibashi}},
  \bibinfo{author}{\bibfnamefont{R.~V.} \bibnamefont{Hijman}},
  \bibinfo{author}{\bibfnamefont{S.}~\bibnamefont{Tarucha}}, \bibnamefont{and}
  \bibinfo{author}{\bibfnamefont{L.~P.} \bibnamefont{Kouwenhoven}},
  \bibinfo{journal}{Nature} \textbf{\bibinfo{volume}{395}},
  \bibinfo{pages}{873} (\bibinfo{year}{1998}).

\bibitem[{\citenamefont{Pakkiam et~al.}(2018)\citenamefont{Pakkiam, Timofeev,
  House, Hogg, Kobayashi, Koch, Rogge, and Simmons}}]{Pakkiam2018}
\bibinfo{author}{\bibfnamefont{P.}~\bibnamefont{Pakkiam}},
  \bibinfo{author}{\bibfnamefont{A.~V.} \bibnamefont{Timofeev}},
  \bibinfo{author}{\bibfnamefont{M.~G.} \bibnamefont{House}},
  \bibinfo{author}{\bibfnamefont{M.~R.} \bibnamefont{Hogg}},
  \bibinfo{author}{\bibfnamefont{T.}~\bibnamefont{Kobayashi}},
  \bibinfo{author}{\bibfnamefont{M.}~\bibnamefont{Koch}},
  \bibinfo{author}{\bibfnamefont{S.}~\bibnamefont{Rogge}}, \bibnamefont{and}
  \bibinfo{author}{\bibfnamefont{M.~Y.} \bibnamefont{Simmons}},
  \bibinfo{journal}{Phys. Rev. X} \textbf{\bibinfo{volume}{8}},
  \bibinfo{pages}{41032} (\bibinfo{year}{2018}).

\bibitem[{\citenamefont{Urdampilleta et~al.}(2019)\citenamefont{Urdampilleta,
  Niegemann, Chanrion, Jadot, Spence, Mortemousque, B{\"{a}}uerle, Hutin,
  Bertrand, Barraud et~al.}}]{Urdampilleta2019}
\bibinfo{author}{\bibfnamefont{M.}~\bibnamefont{Urdampilleta}},
  \bibinfo{author}{\bibfnamefont{D.~J.} \bibnamefont{Niegemann}},
  \bibinfo{author}{\bibfnamefont{E.}~\bibnamefont{Chanrion}},
  \bibinfo{author}{\bibfnamefont{B.}~\bibnamefont{Jadot}},
  \bibinfo{author}{\bibfnamefont{C.}~\bibnamefont{Spence}},
  \bibinfo{author}{\bibfnamefont{P.~A.} \bibnamefont{Mortemousque}},
  \bibinfo{author}{\bibfnamefont{C.}~\bibnamefont{B{\"{a}}uerle}},
  \bibinfo{author}{\bibfnamefont{L.}~\bibnamefont{Hutin}},
  \bibinfo{author}{\bibfnamefont{B.}~\bibnamefont{Bertrand}},
  \bibinfo{author}{\bibfnamefont{S.}~\bibnamefont{Barraud}},
  \bibnamefont{et~al.}, \bibinfo{journal}{Nat. Nanotechnol.}
  \textbf{\bibinfo{volume}{14}}, \bibinfo{pages}{737} (\bibinfo{year}{2019}).

\bibitem[{\citenamefont{West et~al.}(2019)\citenamefont{West, Hensen, Jouan,
  Tanttu, Yang, Rossi, Gonzalez-Zalba, Hudson, Morello, Reilly
  et~al.}}]{West2019}
\bibinfo{author}{\bibfnamefont{A.}~\bibnamefont{West}},
  \bibinfo{author}{\bibfnamefont{B.}~\bibnamefont{Hensen}},
  \bibinfo{author}{\bibfnamefont{A.}~\bibnamefont{Jouan}},
  \bibinfo{author}{\bibfnamefont{T.}~\bibnamefont{Tanttu}},
  \bibinfo{author}{\bibfnamefont{C.~H.} \bibnamefont{Yang}},
  \bibinfo{author}{\bibfnamefont{A.}~\bibnamefont{Rossi}},
  \bibinfo{author}{\bibfnamefont{M.~F.} \bibnamefont{Gonzalez-Zalba}},
  \bibinfo{author}{\bibfnamefont{F.}~\bibnamefont{Hudson}},
  \bibinfo{author}{\bibfnamefont{A.}~\bibnamefont{Morello}},
  \bibinfo{author}{\bibfnamefont{D.~J.} \bibnamefont{Reilly}},
  \bibnamefont{et~al.}, \bibinfo{journal}{Nat. Nanotechnol.}
  \textbf{\bibinfo{volume}{14}}, \bibinfo{pages}{437} (\bibinfo{year}{2019}).

\bibitem[{\citenamefont{Zheng et~al.}(2019)\citenamefont{Zheng, Samkharadze,
  Noordam, Kalhor, Brousse, Sammak, Scappucci, and Vandersypen}}]{Zheng2019}
\bibinfo{author}{\bibfnamefont{G.}~\bibnamefont{Zheng}},
  \bibinfo{author}{\bibfnamefont{N.}~\bibnamefont{Samkharadze}},
  \bibinfo{author}{\bibfnamefont{M.~L.} \bibnamefont{Noordam}},
  \bibinfo{author}{\bibfnamefont{N.}~\bibnamefont{Kalhor}},
  \bibinfo{author}{\bibfnamefont{D.}~\bibnamefont{Brousse}},
  \bibinfo{author}{\bibfnamefont{A.}~\bibnamefont{Sammak}},
  \bibinfo{author}{\bibfnamefont{G.}~\bibnamefont{Scappucci}},
  \bibnamefont{and} \bibinfo{author}{\bibfnamefont{L.~M.~K.}
  \bibnamefont{Vandersypen}}, \bibinfo{journal}{Nat. Nanotechnol.}
  \textbf{\bibinfo{volume}{14}} (\bibinfo{year}{2019}).

\bibitem[{\citenamefont{Vu and {Das Sarma}}(2020{\natexlab{a}})}]{Vu2020}
\bibinfo{author}{\bibfnamefont{D.}~\bibnamefont{Vu}} \bibnamefont{and}
  \bibinfo{author}{\bibfnamefont{S.}~\bibnamefont{{Das Sarma}}},
  \bibinfo{journal}{Phys. Rev. B} \textbf{\bibinfo{volume}{101}},
  \bibinfo{pages}{125113} (\bibinfo{year}{2020}{\natexlab{a}}).

\bibitem[{\citenamefont{Vu and {Das Sarma}}(2020{\natexlab{b}})}]{Vu2020b}
\bibinfo{author}{\bibfnamefont{D.}~\bibnamefont{Vu}} \bibnamefont{and}
  \bibinfo{author}{\bibfnamefont{S.}~\bibnamefont{{Das Sarma}}},
  \bibinfo{journal}{Phys. Rev. Res.} \textbf{\bibinfo{volume}{2}},
  \bibinfo{pages}{023060} (\bibinfo{year}{2020}{\natexlab{b}}).

\bibitem[{\citenamefont{Townsend et~al.}(2021)\citenamefont{Townsend, Neuman,
  Debrecht, Aizpurua, and Bryant}}]{Townsend2021}
\bibinfo{author}{\bibfnamefont{E.}~\bibnamefont{Townsend}},
  \bibinfo{author}{\bibfnamefont{T.}~\bibnamefont{Neuman}},
  \bibinfo{author}{\bibfnamefont{A.}~\bibnamefont{Debrecht}},
  \bibinfo{author}{\bibfnamefont{J.}~\bibnamefont{Aizpurua}}, \bibnamefont{and}
  \bibinfo{author}{\bibfnamefont{G.~W.} \bibnamefont{Bryant}},
  \bibinfo{journal}{Phys. Rev. B} \textbf{\bibinfo{volume}{103}},
  \bibinfo{pages}{195429} (\bibinfo{year}{2021}).

\end{thebibliography}

\appendix

\section{Full set of charge-stability diagrams \label{app:csd_full}}
\begin{figure*}
   \centering
    \includegraphics[width=\textwidth]{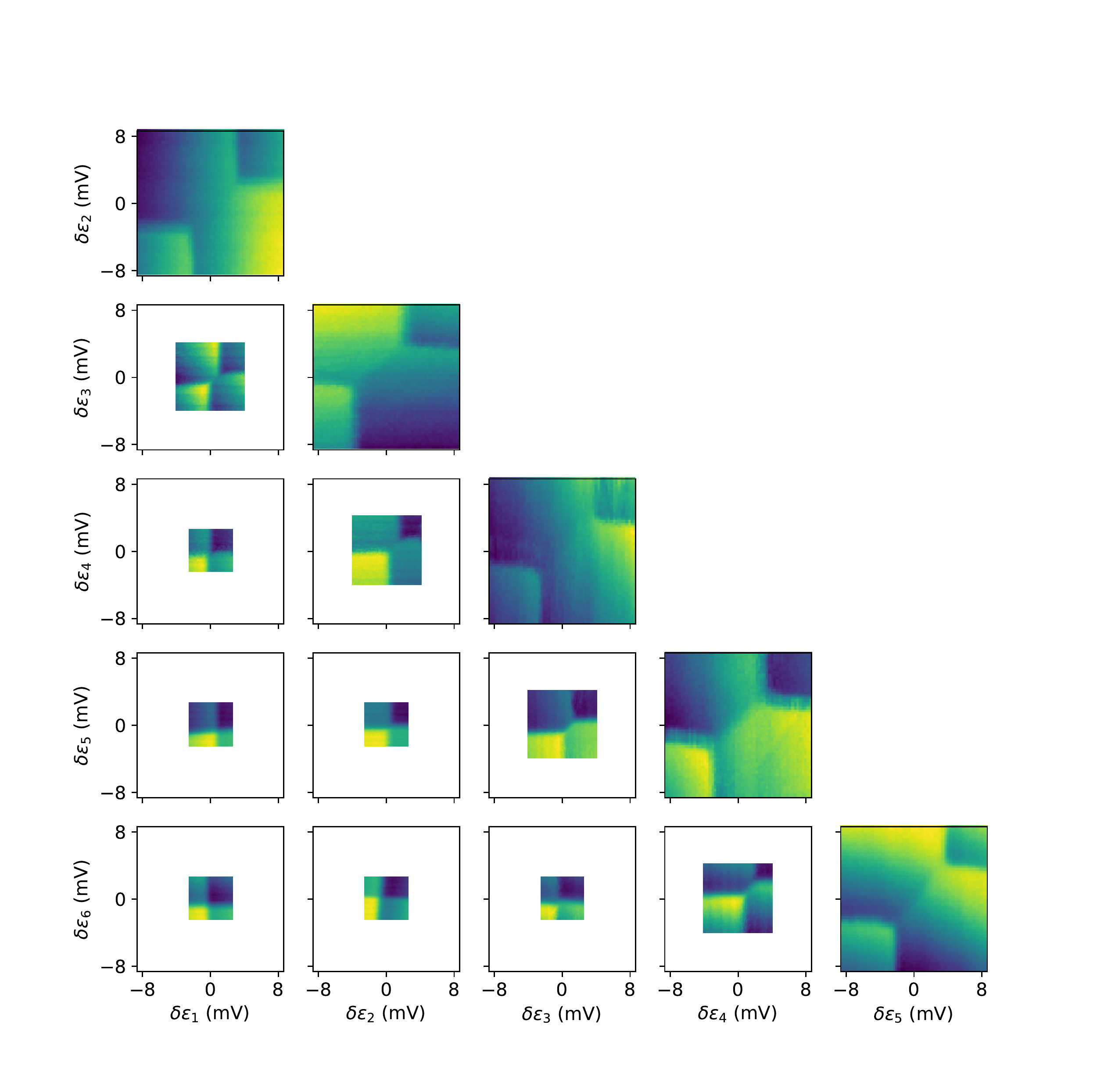}
   \caption{Full set of pairwise charge-stability diagrams. The leftmost column shows the diagrams presented in Fig.~\ref{fig:int_exp} in the main text. For non-nearest neighbour pairs smaller voltage ranges were used to maintain sufficient resolution with the same number of data points.}
   \label{fig:csd_full}
\end{figure*}

In the main text, only the charge-stability diagrams involving the leftmost dot are provided.
The full set of pairwise charge-stability diagrams is shown in Fig.~\ref{fig:csd_full}.

\section{Screened interaction potential \label{app:screening}}
In this Appendix, we summarize our theoretical analysis to estimate the interaction strength as a function of distance, as depicted in Fig.~\ref{fig:int_exp} of the main text.
The main task is to derive a form of the interaction potential $V_\mathrm{ee}$.
Given that potential, the matrix elements $V_{ij}$ and $U_i = V_{ii}$ can be obtained within a Wannier basis $\{ \phi_i \}_{i=1, ..., N}$ as
\begin{equation}\label{eq:matrix-elements}
V_{ij} = \int \mathrm{d}^2 \mathbf{r} \int \mathrm{d}^2 \mathbf{r}^\prime |\phi_i(\mathbf{r})|^2 V_\mathrm{ee}(\mathbf{r},\mathbf{r}^\prime) |\phi_j(\mathbf{r)^\prime}|^2,
\end{equation}
where $V_\mathrm{ee}$ denotes the two-body interaction potential between electrons.
These elements can be checked for self-consistency by comparison with the tunnel coupling elements given by
\begin{equation}
t_{ij} = \int \mathrm{d}^2 \mathbf{r} \phi_i^*(\mathbf{r}) [ - \frac{\hbar^2}{2m} \nabla^2 + V(\mathbf{r}) ] \phi_j(\mathbf{r}).
\end{equation}
The basis states $\phi_i$ are constructed from Gaussians centered around the central locations of the QDs, denoted by $\mathbf{\tau}_i$.

\subsection{Method of image charges}
\begin{figure}
   \centering
    \includegraphics[width=\columnwidth]{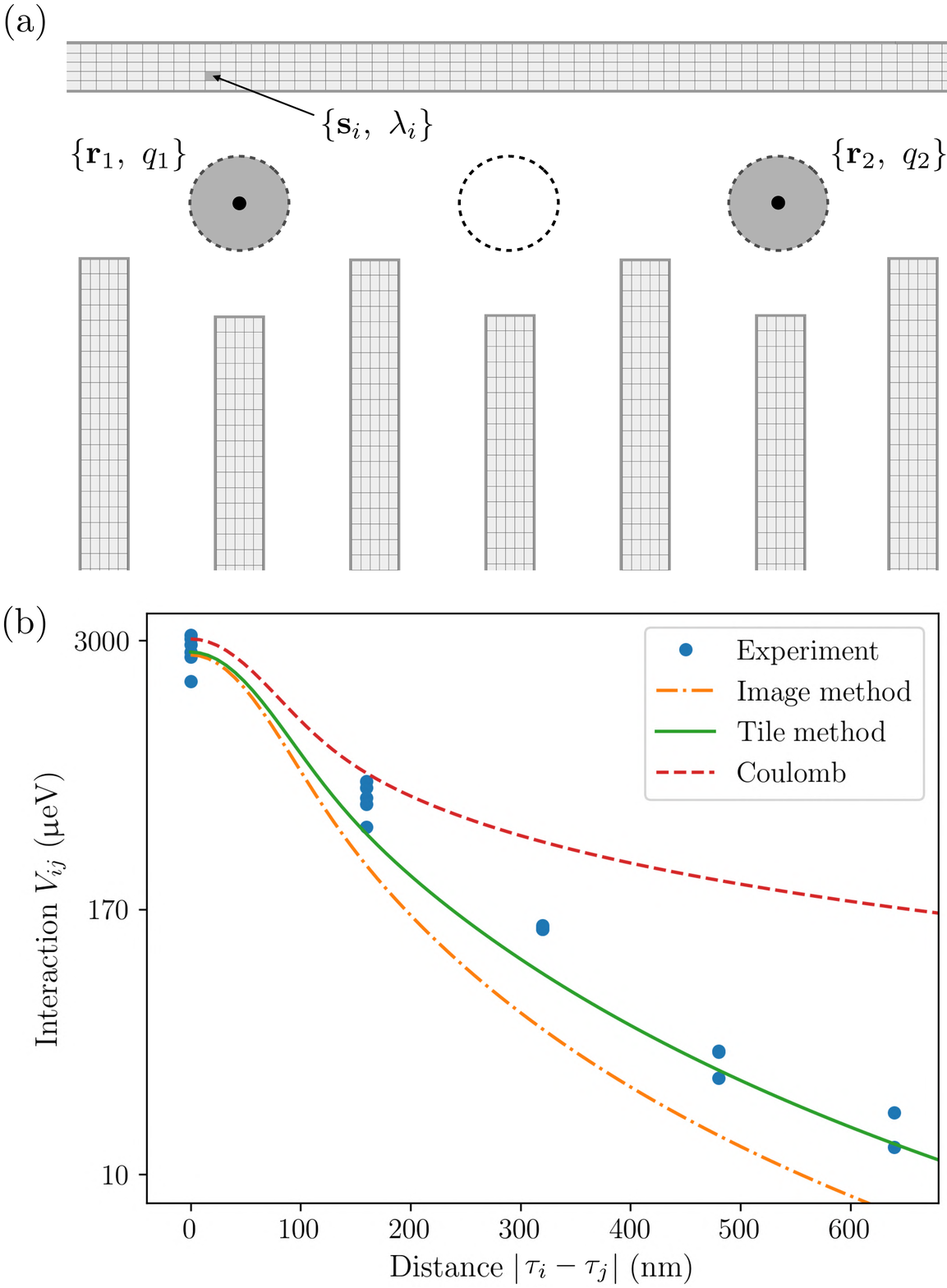}
   \caption{
            Screened interaction potential.
			(a) Schematic illustration of numerical procedure to calculate the screened interaction potential.
			The device geometry shown in Fig.~\ref{fig:int_exp} is taken and its surface area discretized.
			The interaction potential is obtained from the pairwise interaction terms of all tile charges $\lambda_i$ at position $\mathbf{s}_i$
			and the two electrons at $\mathbf{r}_1$ and $\mathbf{r}_2$.
			(b) Interaction matrix elements $V_{ij}$, between dots located at $\mathrm{\tau}_i$ and $\mathbf{\tau}_j$, compared for different cases:
			unscreened Coulomb interactions (\textit{red, dashed}), screened interactions via image-charge method for a surface completely covered by metal (\textit{orange, dash-dotted})
			and screened interaction via numerical discretization of real device geometry (\textit{green, solid}).
			}
   \label{fig:app_screening}
\end{figure}

We consider screening due to metallic gates as the dominant screening source in our sample.
For a conservative estimate, we first calculate the screened Coulomb potential under the assumption that the whole surface layer was covered by metal.
Using the method of image charges, the screened interaction between two electrons at $\mathbf{r}_1 = (x_1, y_1, -d)$ and $\mathbf{r}_2 = (x_2, y_2, -d)$ takes the form
\begin{equation}\label{eq:app_image-charge}
\begin{aligned}
f_\mathrm{im}(\mathbf{r}_1, \mathbf{r}_2) = & \ 1 - \frac{\sqrt{(x_1-x_2)^2 + (y_1-y_2)^2}}{\sqrt{(x_1-x_2)^2 + (y_1-y_2)^2 + 4d^2}},\\
V_\mathrm{ee}^\mathrm{im}(\mathbf{r}_1, \mathbf{r}_2) = & \ f_\mathrm{im}(\mathbf{r}_1, \mathbf{r}_2) \frac{k_0 e^2}{|\mathbf{r}_1 - \mathbf{r}_2|}\\
& \xrightarrow{d/|\mathbf{r}_1-\mathbf{r}_2|\rightarrow0}\frac{2k_0(ed)^2}{|\mathbf{r}_1-\mathbf{r}_2|^3},
\end{aligned}
\end{equation}\newline
with $k_0 = 1/(4\pi\varepsilon\varepsilon_0)$.
In our numerical simulations we use $\varepsilon = 12.9$, the relative dielectric constant for GaAs.
As Fig.~\ref{fig:app_screening}(b) shows, this ansatz overestimates the screening effect as expected.
In our sample, the surface area is only partially covered by thin metallic gates.
In the following Sec.~\ref{ssec:charge-tiling}, we calculate the screened potential due to the gates in the real device geometry.

\subsection{Charge-tiling method \label{ssec:charge-tiling}}
To calculate the metal-induced screening numerically, we start from the real device geometry and discretize its surface area into $m$ tile charges $\lambda_i~(i = 1, ..., m)$ centered around positions $\mathbf{s}_i$, cf.~Fig.~\ref{fig:app_screening}(a).
We derive the interaction $V_\mathrm{ee}(\mathbf{r}_1, \mathbf{r}_2)$ for two electrons located $d = 90~\mathrm{nm}$ below the surface at $\mathbf{r}_1$ and $\mathbf{r}_2$, respectively.
Here, $\{\lambda_i\}_{i=1, ..., m}$ refers to the induced tile charges due to the presence of these two electrons.
With the electron charge $q_1 = q_2 = -e$, we obtain the electrostatic potential at tile $i$ as
\begin{equation}
    \tilde V(\mathbf{s}_i) = \sum_{k=1,2} \frac{-k_0e}{|\mathbf{s}_i - \mathbf{r}_k|} + \sum_{j\neq i} \frac{k_0\lambda_j}{|\mathbf{s}_j-\mathbf{s}_i|}.
\end{equation}
We assume charge conservation in the metallic layer, $\sum_i \lambda_i = 0$, and a constant potential, $\tilde V(\mathbf{s}_i) = \tilde V(\mathbf{s}_j), \forall i, j$.
This yields a system of linear equations that can be solved to determine the tile charges $\lambda_i$.
Finally, the screened interaction potential energy can be obtained from
\begin{equation}\label{eq:app_screened_potential}
\begin{aligned}
V_\mathrm{ee}(\mathbf{r}_1, \mathbf{r}_2) =& \frac{-e}{2} \sum_{k=1,2} \sum_{i=1}^m \frac{k_0\lambda_i}{|\mathbf{r}_k-\mathbf{s}_i|} \\
& + \frac{k_0e^2}{|\mathbf{r}_1-\mathbf{r}_2|}.
\end{aligned}
\end{equation}

The first term of the right-hand side of Eq.~\eqref{eq:app_screened_potential} shows that the tile charges screen the bare Coulomb interaction $\sim 1/|\mathbf{r}_1-\mathbf{r}_2|$.
This is demonstrated in Fig.~\ref{fig:app_screening}(b).
The numerically obtained results lie between the bare Coulomb and fully screened (based on Eq.~\eqref{eq:app_image-charge}) curves.
They are in good agreement with the measurements outcomes and demonstrate the long-range character of $V_\mathrm{ee}$.
We note that the size of the quantum dots, as described by the FWHM of the functions $\phi_i$, is a fitting parameter in our numerical approach.
We find this size to be $\approx 45$nm.

\section{Charge occupation numbers \label{app:occ_num_mol}}
\begin{figure*}
   \centering
    \includegraphics[width=0.7\textwidth]{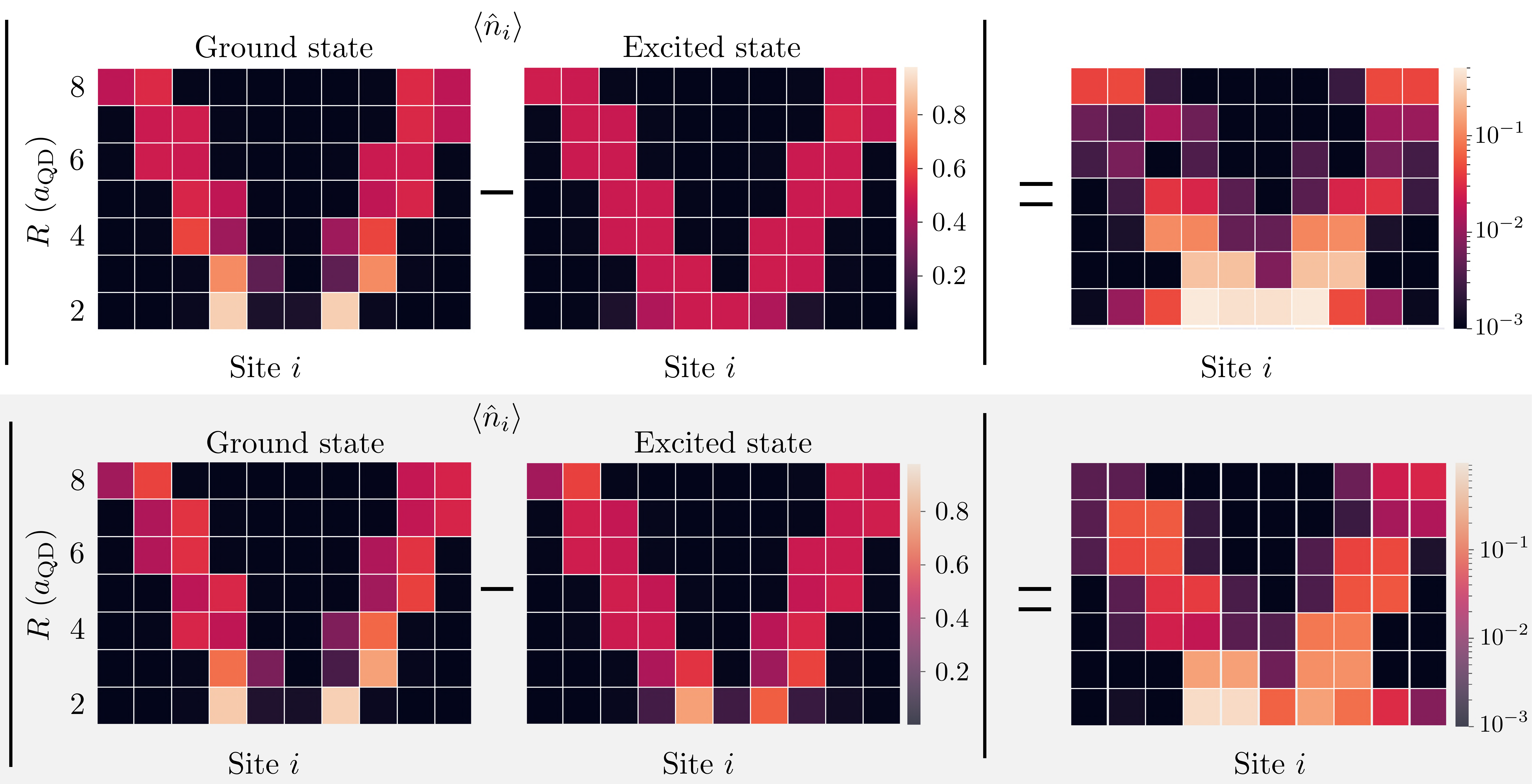}
   \caption{
            Charge occupation numbers $\braket{\hat n_i}$ for a system with $N = 10$ QDs and two electrons.
            \textit{Upper panel:} Results for ground (left) and first excited (middle) state, respectively.
            The $x$ axis denotes the site index of the $i$th dot in the array, and the $y$ axis denotes the internuclear separation $R/a_\mathrm{QD}$.
            The absolute value of the difference between the first two columns is shown on the right.
            \textit{Lower panel:}
            Same as upper panel, but for a model with an additional bias term $\Delta \varepsilon_i = (i - x_\mathrm{c}) \cdot \SI{10}{\micro\relax}$eV, where $x_\mathrm{c}$ denotes the center of the QD array.
			}
   \label{fig:app_occupation}
\end{figure*}

The expectation values $\braket{\hat n_i}$ describe the spread of the electronic wavefunction over the QD array, cf.~Fig.~\ref{fig:schematic}.
As described in the main text in Sec.~\ref{ssec:experimental-techniques}, resonant microwave excitations may be employed for probing transitions from ground to low-lying excited states.
In Fig.~\ref{fig:app_occupation} we show $\braket{\hat n_i}$ for an artificial two-electron molecule, both for the ground and first-excited states.
As shown in the figure, the difference in local occupation between ground and excited state may be enhanced by applying a potential bias, e.g., in the form of $\varepsilon_i \propto i$.

\end{document}